\documentclass{article}

\usepackage{arxiv}

\usepackage[utf8]{inputenc} 
\usepackage[T1]{fontenc}    
\usepackage{hyperref}       
\usepackage{url}            
\usepackage{booktabs}       
\usepackage{amsmath}
\usepackage{amsfonts}       
\usepackage{nicefrac}       
\usepackage{microtype}      
\usepackage{lipsum}
\usepackage{fancyhdr}       
\usepackage{graphicx}       
\graphicspath{{media/}}     

\title{Fundamental of CO$_2$ Adsorption and Diffusion in Sub-nanoporous Materials: Application to CALF-20}

\author{
  Andr\'e de Freitas Gon\c calves\\
  Universidade Estadual de Campinas (UNICAMP), Faculdade de Engenharia Química, Campinas, CEP:13083-852, SP, Brazil.\\
  \AND
  Emerson Parazzi Lyra\\
  Universidade Estadual de Campinas (UNICAMP), Faculdade de Engenharia Química, Campinas, CEP:13083-852, SP, Brazil.\\
  \AND
  Sayali Ramdas Chavan\\
  TotalEnergies, OneTech, Pole d’Etudes et de Recherche de Lacq (PERL), Lacq, 64170, France.\\
  \AND
  Philip L. Llewellyn\\
  TotalEnergies, OneTech, Sustainability R\&D, CSTJF, Pau, 64018, France.\\
  \AND
  Luis Fernando Mercier Franco\\
  Universidade Estadual de Campinas (UNICAMP), Faculdade de Engenharia Química, Campinas, CEP:13083-852, SP, Brazil.\\
  \AND
  Yann Magnin\\
  TotalEnergies, OneTech, R\&D, CSTJF, Pau, 64018, France.\\
  \textit{yann.magnin@totalenergies.com} \\
}

\begin{document}
\maketitle

\begin{abstract}
We propose a theoretical approach for predicting thermodynamics and kinetics of guest molecules in nanoporous materials. This statistical mechanical-based method requires a minimal set of physical parameters that may originate from experiments or numerical simulations. We applied it to CO$_2$ molecules in the recently highlighted CALF-20 metal-organic framework for adsorption and molecular self-diffusion at different temperatures. All the physical parameters of the model are extracted from one CO$_2$ isotherm analyzed by the adsorption energy distribution method. The model is then used to approximate isotherms at different temperatures, Henry's constant, saturation density, as well as enthalpies of adsorption at infinite dilution. We then express molecular kinetics through the transition state theory allowing to predict molecular diffusion in part from the prior knowledge of thermodynamics, and further compared self-diffusion coefficients to one from molecular dynamics used as a numerical experiment. The approach proposed allows to express molecular adsorption and diffusion based on a fitting procedure allowing to get physical parameters with a view on thermodynamics and kinetics mechanisms at play in the system.
\end{abstract}

\bigskip

Carbon capture, transport and storage is one key approach to carbon mitigation. In this value chain, the capture step is often considered the most onerous.\cite{koutsonikolas2016} Significant research and development is thus devoted to optimizing existing approaches and promoting second generation technologies.\cite{kearns2021} Amine scrubbing is considered as one of the key and mature approach where CO$_2$ molecules chemisorb in solution with quite rapid capture rate and relatively high enthalpies.\cite{orlov2022} A possible alternative consists of adsorption on porous materials, often via physisorption, allowing lower enthalpies of adsorption offering theoretically a less expensive regeneration compared to chemisorption.\cite{gao2023} Although physisorption is promising, it also presents bottlenecks and opens new challenges. A first is the choice of active materials in the wide zoology of porous solids spanning through zeolites,\cite{perez2022} activated carbons,\cite{sircar1996} carbon molecular sieves,\cite{reid1998} metal-organic frameworks (MOF),\cite{li2009,wang2017} silica based materials,\cite{dutcher2015} porous polymers,\cite{dawson2012} covalent-organic framework,\cite{zeng2016} among others. One important constraint is the ability of such materials to remain stable and maintain a reasonable CO$_2$ capacity after a substantial number of capture cycles. An-other is to account for the CO$_2$ interactions with various contaminants in the flue gases, and particularly humidity, which are unavoidable in most carbon capture scenarios, whether these be during post-combustion or, to a lesser extent, direct air capture.\cite{kumar2015} While CO$_2$/H$_2$O equilibrium measurements have recently been highlighted,\cite{nguyen2024_c1} there is still space for novel and less time-consuming methods to enable systematic studies.\cite{rajendran2024_2,polat2024,chanut2017,kolle2021}\\
To overcome some of the challenges to physisorption including productivity, rapid capture approaches have been proposed. Within it, physisorption is performed with minute-long cycles in both rapid-PSA (pressure swing adsorption) and rapid-TSA (temperature swing adsorption) processes.\cite{subramanian2019,subraveti2019} These techniques are evaluated from thermodynamics models allowing to predict the process working capacity. However the models currently used are often empirical or semi-empirical, and chosen from their ability to fit experimental measurements.\cite{majd2022} While these latter have been instrumental in engineering, they do not allow to give clear insights into the adsorption mechanisms at play. Even those based on statistical mechanical arguments,\cite{franco2017,araujo2019} or within classical Density Functional Theory framework,\cite{goncalves2024,bardow2025} may over- or underestimate some parameters due to the fitting procedure falling into local minimums of the parameter space. Fitted parameters may not be transferable from one temperature to an other in theory and some adsorption mechanisms may be missed when the fitting procedure is applied in a limited range of pressure.\\
Beyond equilibrium thermodynamics, diffusion of guests molecules, and more precisely mass transfer are important parameters in areas such as carbon capture,\cite{alessandro2010,bui2018,paltsev2021,choi2009} where this determines the process production rate.\cite{verstreken2024} Typically, processes operate through shaped porous sorbents consisting of nanoporous active materials (such as MOFs) embedded into binders forming a multi-scale porous material with porosity ranging from the sub-nanometer (nanoporosity, \textit{i.e.} $<$ 1 nm) to hundred of nanometers (macroporosity, \textit{i.e.} $>$ 50 nm).\cite{rouquerol1994,verstreken2024} In these solids, various mass transfer resistances applied at different length and time scales. In a fine tuning of the capture process, it can thus be of interest to identify and optimize the limiting diffusion phenomena to optimize the process productivity.\cite{verstreken2024}\\
Whilst the selection of active materials should depend on both thermodynamic and kinetic properties, it would seem that kinetics have received less attention, in part due to the challenge in accurately measuring kinetics properties experimentally.\cite{verstreken2024,grossmann2025} Especially considering the project of real process units for carbon capture, it is crucial to have reliable and consistent models for the equipment sizing and for the economic evaluation of the process. Hence, this study aims at proposing a predictive physical model, based on a fitting procedure allowing determination of physical parameters to describe thermodynamics and kinetics, here limited to CO$_2$ molecules in the sub-nanoporosity of the recent CALF-20 MOF,\cite{lin2021} here chosen as a benchmark. Technical details concerning the numerical simulations, data treatments and experiments are provided in the supporting information, also including the AED source code used in this work. In practice, this simple general model may be easily used as proposed in other solids, and we hope it may be of interest for engineering approaches dealing with nanoporous materials.

\section{CALF-20 Metal-Organic Framework}
\begin{figure}[h!]
\centering
\includegraphics[scale=0.22]{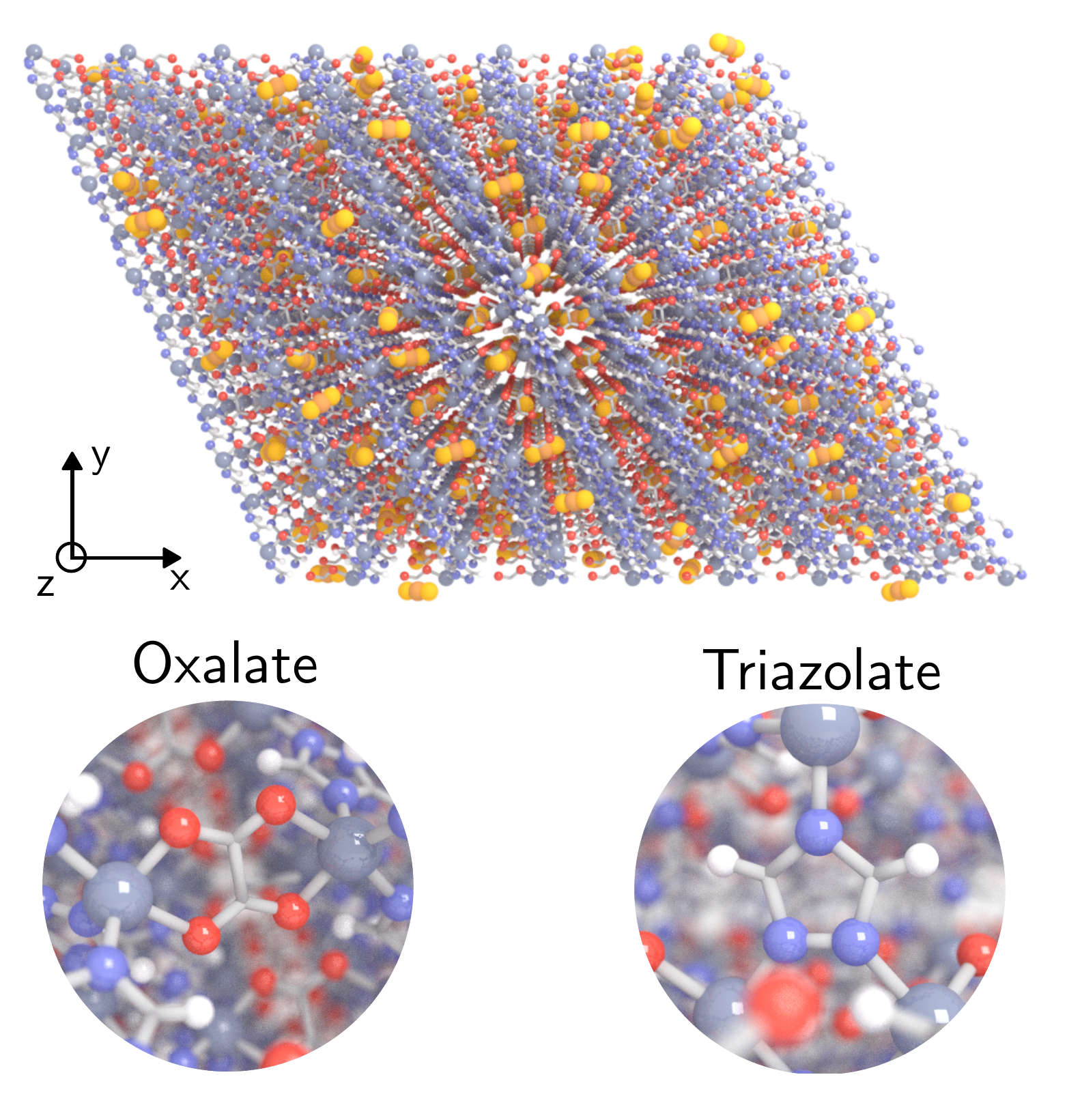}
\caption{CALF-20 structure made of oxalate and triazolate linkers bonding zinc metal atoms. Adsorbed CO$_2$ molecules (yellow molecules) are shown at $P$=0.25 bar and $T$=293.15 K.}
\label{fig:figure1}
\end{figure}

CALF-20(Zn) is a sub-nanoporous MOF structure initially reported by Shimizu \emph{et al.} in 2021\cite{lin2021} Figure.\ref{fig:figure1}A. This material made of cages connected by narrow channels Figure.\ref{fig:figure1}B has shown a tremendous interest thanks to relatively rare properties in regard to the MOF structural space such as, its stability to humidity and harsher contaminants, such as NO$_x$ and SO$_x$, its ability to capture CO$_2$ from wet flue gases, and the possibility of being prepared via a rather cheap and scalable synthesis procedure. All together, these attributes make CALF-20 a serious alternative sorbent to the 13X zeolite, considered as a target reference in terms of CO$_2$ capacity, while this latter remains of limited practical interest due to its large hydrophilicity.\cite{chen2014,dhoke2021} The company Svante has already demonstrated the efficiency of the CALF-20 by integrating it in its rapid temperature swing adsorption process, shown to capture up to 1 tonne of CO$_2$ per day from cement plant flue gases.\cite{hovington2022} At the atomistic scale, the CALF-20 structure is made of 1,2,4-triazolate-bridged zinc(II) ions layers pillared by oxalate ions linkers to form a three-dimensional lattice with a pore size of $\sim$ 0.7 nm, Figure.\ref{fig:figure1}. Further details concerning the CALF-20 MOF structure, the different CALF-20 phases, flexibility of the structure, thermal conductivity, adsorption and diffusion mechanisms of CO$_2$, H$_2$O as well as mixtures can be found in the recent CALF-20 literature.\cite{lin2021,nguyen2022,ymagnin2023,nguyen2023,chen2023,naskar2023,ho2023,gopalsamy2024,rajendran2024,wang2024,fan2024,oktavian2024,hasting2024,krishna2024,drwkeska2024,lassitter2024,drwkeska2025,nguyen2024_c1,attallah2025,pereira2025,constant2025,duplessis2025,dilipkumar2025,dilipkumar2025_2}

\section{Thermodynamics}
\subsection{CO$_2$ adsorption in CALF-20(Zn) by atomistic simulations}
\begin{figure}[h!]
\centering
\includegraphics[scale=0.25]{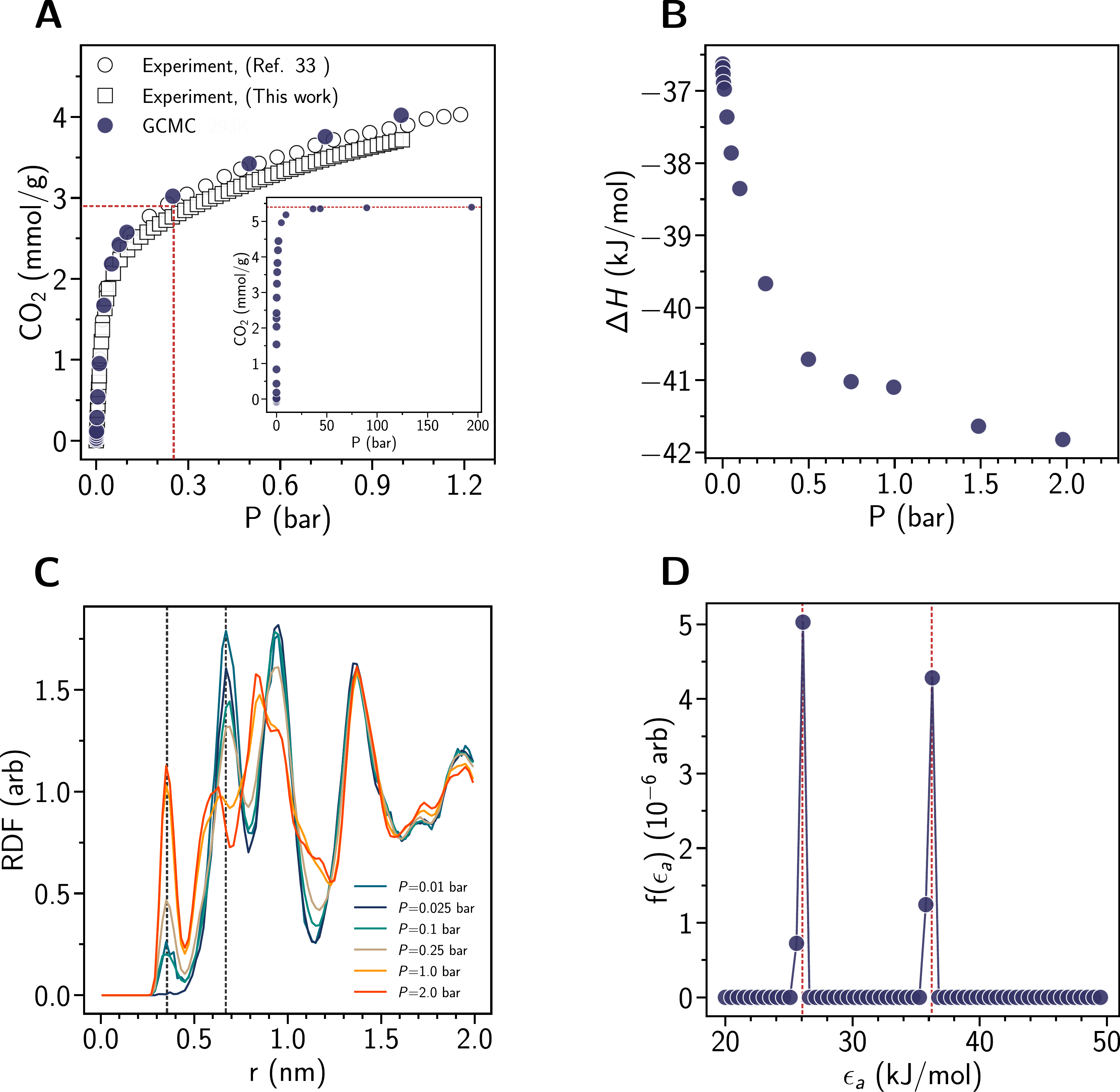}
\caption{A. CO$_2$ adsorption isotherm in CALF-20 at $T$=293.15K. The blue circles correspond to GCMC simulations, the open symbols to experiments made in this work (squares) and from the literature (circles).\cite{lin2021} The red dashed line indicate the loading corresponding to $P$=0.25 bar. The inset shows the isotherm over a large pressure range and the dashed red line denotes the saturation density. B. Enthalpy of adsorption by GCMC simulations at $T$=293.15K. C. Radial distribution function of CO$_2$ molecules from their center of mass for different pressures at $T$=293.15K. The dashed gray lines correspond to the first and second adsorption sites into the CALF-20 pores. D. Adsorption energy distribution applied to the full $P$ range in GCMC isotherm (A). The red dashed line denotes the CO$_2$ adsorption energies on the different adsorption sites.}
\label{fig:figure2}
\end{figure}

The CO$_2$ adsorption isotherm in CALF-20 by grand canonical Monte Carlo simulations (GCMC) at $T$=293.15K are shown in the Figure.\ref{fig:figure2}A. Simulations (blue circles) and experiments (open circles and open squares)\cite{lin2021} are found to be in good agreement, all presenting a Langmuir shape, characteristic of nanoporous adsorbents (experimental details are provided in supporting information). The enthalpy of adsorption, determined from the fluctuation method,\cite{vuong1996} was shown to increase (more negative values) with the loading, Figure.\ref{fig:figure2}B, in agreement with experimental results already reported in the literature.\cite{lin2021,dilipkumar2025_2} At moderate pressure, single CO$_2$ are adsorbed per CALF-20 cages.\cite{lin2021,drwkeska2024} The increasing behavior of the enthalpy has been attributed to the increasing number of guest interacting from a cage to another.\cite{ymagnin2023} Thus, from Figure.\ref{fig:figure2}B, we can notice that collective guests interactions are responsible from about $\sim$9\% increase in $\Delta H$ from infinite dilution to $P$=2 bar, and remains almost independent of the temperature at given loadings, (SI.Figure.1).

\subsection{CO$_2$ adsorption in CALF-20(Zn) by statistical mechanics}
\begin{figure}[h!]
\centering
\includegraphics[scale=0.4]{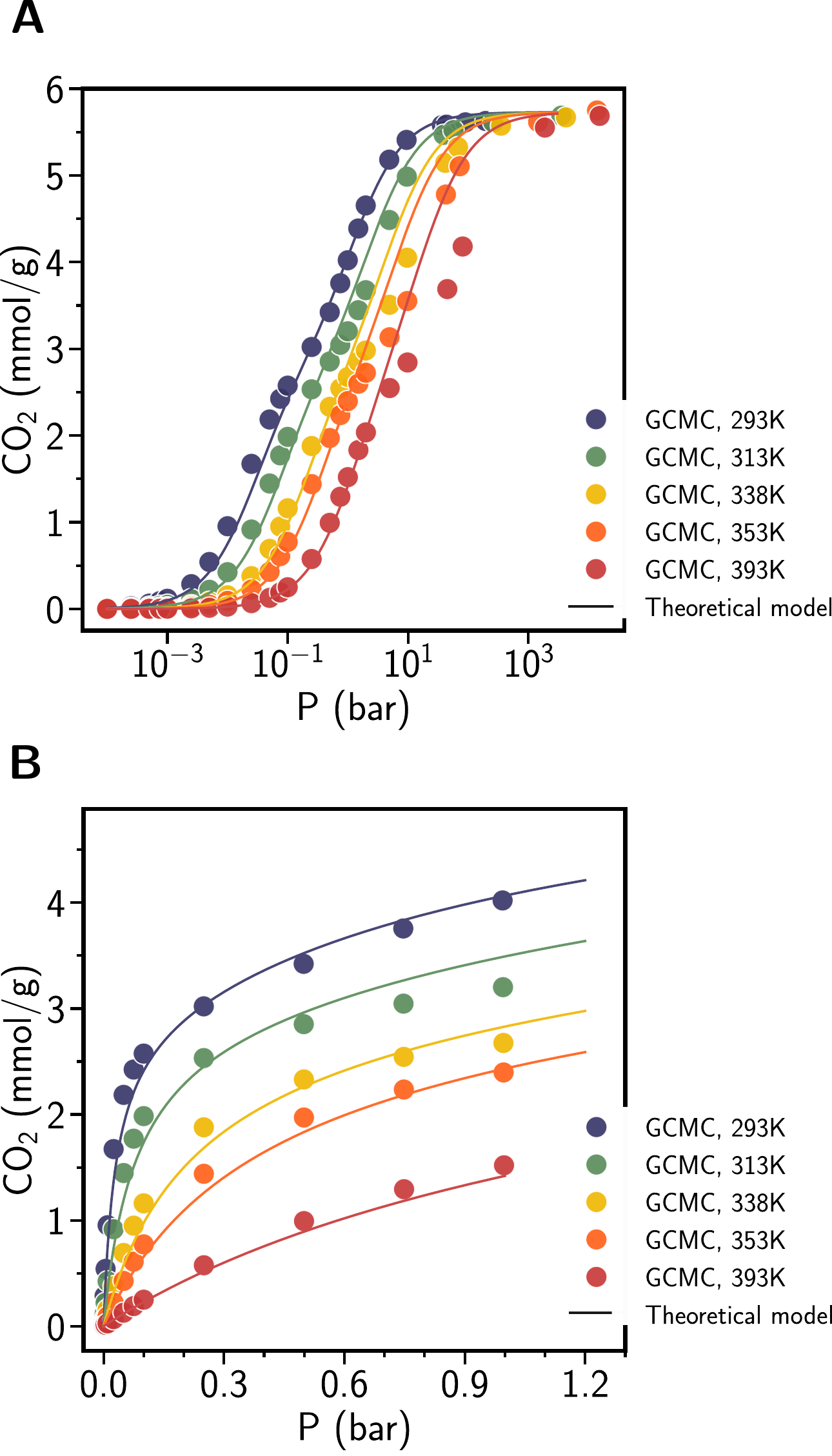}
\caption{A. CO$_2$ isotherms (lin-log plot) by GCMC (colored circles) and theoretical model (colored lines). Isotherms are shown at different temperatures, $T$=293.15K (blue circles), 313.15K (green circles), 338.15K (yellow circles), 353.15K (orange circles) and $T$=393.15K (red circles) on a large $P$ range. B. Same than (A), on a limited range of pressure.}
\label{fig:figure3} 
\end{figure}

\noindent While such equilibrium parameters are today commonly measured experimentally and/or determined from simulations,\cite{rouquerol2013} getting this in the early 20th century was far from trivial. Hence, in 1918, Langmuir proposed a model to describe the thermodynamic equilibrium of an ideal gas reservoir interacting with a solid surface under isothermal and isobaric conditions (constant $P$ and $T$).\cite{langmuir1918} According to this model, identical adsorption sites are considered to be homogeneously distributed at the surface of a perfectly flat solid. The guest-host binding is considered identical from one site to an other and independent of both the temperature and guest loading. Each adsorbate can only occupy one site and are approximated as static (no molecular diffusion or vibration induced by the thermal excitation). Finally, this model neglects lateral interactions in between adsorbed molecules forming a single layer adsorption onto the solid surface. The single site Langmuir's equation is commonly expressed such as,
\begin{eqnarray}
\label{eq:langmuir_orig}
\theta &=& \frac{\rho_a}{\rho_s}\ = \ \frac{a(T)\times P}{1\ +\ a(T)\times P},
\end{eqnarray}
with $\theta$ the adsorbate coverage corresponding to the ratio of the adsorbate density $\rho_a$ (at given $P$ and $T$) and the saturation density $\rho_s$, corresponding to the full occupation of adsorption sites in the solid. In the last term of \eqref{eq:langmuir_orig}, $a(T)$ is an equilibrium constant (at a given $T$), usually fitted from experiments, lumping together parameters such as the vapor pressure, the temperature, the binding energy, \textit{etc}, in other words, all the  physico-chemical complexity of the system. It is possible to reformulate \eqref{eq:langmuir_orig} in a more detailed manner by deriving it from the statistical mechanics. To do so, we write the two partition functions $Z_g$ and $Z_a$ corresponding to the gas phase and adsorbed molecules, respectively. The gas partition function can be easily written such as,\cite{hill1986}
\begin{eqnarray}
\label{eq:Zg}
\mathsf{Z}_g & = & \frac{1}{n_g!}\left\lbrace\left(\frac{2\pi m k_BT}{h^2}\right)^{3/2}\  V\right\rbrace^{n_g},
\end{eqnarray}
considering indistinguishable and non interacting CO$_2$ molecules (ideal gas), with $k_B$ the Boltzmann constant, $m$ the mass of a CO$_2$ molecule, $h$ the Planck constant and $V=n_g k_BT/P$ the volume of the ideal gas reservoir, with $n_g$ the number of molecules populating it. The factor $1/n_g!$ comes from the indistinguishability of gas molecules, normalizing the over summation of similar classical microstates. In \eqref{eq:Zg}, we can also notice that we neglect the rotational and vibrational molecular partition function (the first being $\sim$ 1 and the second being neglected in numerical simulations) for a sake of simplicity, restricting the approach to the three dimensional translation partition function expressed by the de Broglie thermal wavelength, $\Lambda=h/(2\pi m k_BT)^{1/2}$. The partition function corresponding to the adsorbed phase writes,
\begin{eqnarray}
\label{eq:Za}
\mathsf{Z}_{a} &=& \frac{N!}{n_a!(N-n_a)!}\left\lbrace\left[2\sinh\left(\frac{h\nu_a}{2k_BT}\right)\right]^{-3}\ \exp\left(\frac{\epsilon^0_a}{k_BT}\right)\right\rbrace^{n_a},
\end{eqnarray} 
with the combinatorial term corresponding to a normalization factor denoting the number of possible configurations to distribute $n_a$ indistinguishable adsorbates on $N$ distinguishable adsorption sites, with $(N-n_a)$ the number of vacant adsorption sites. The first term in the brackets corresponds to the Einstein's crystal vibration partition function,\cite{hill1986} denoting independent harmonic oscillators, all approximated with a frequency $\nu_a$, corresponding to the oscillations of the adsorbed molecules on their respective adsorption sites. The second term (in the brackets) is related to the adsorption probability onto the surface at a given $T$, with $\epsilon^0_a$ (defined as a positive value in our calculations) denoting the guest-hots binding energy at infinite dilution and $T$= 0K. As mentioned above, the Langmuir equation expresses the thermodynamic equilibrium lying in between a surface covered by adsorbates with a surrounding gas phase. In such a case, chemical potentials of the adsorbed phase and the gas reservoir are equivalent, \textit{i.e.} $\mu_a=\mu_g$. From \eqref{eq:Zg} and \eqref{eq:Za} we can express these chemical potentials from $\mu=-k_BT\ \partial Z/ \partial n$ using Stirling's approximation stating that $\ln(n!) \sim n\ln(n)-n$ for $n\gg$1,
\begin{eqnarray}
 \label{eq:mug}
\mu_g &=&\ k_BT\ \ln \left\lbrace\left(\frac{h^2}{2\pi m k_BT}\right)^{3/2}\ \frac{P}{k_BT}\right\rbrace,\\
\label{eq:mua}
\mu_{a} &=& k_BT \ln \left\lbrace \frac{\theta}{1-\theta}\ \left[2\sinh\left(\frac{h\nu_a}{2k_BT}\right)\right]^{3}\ \exp\left(-\frac{\epsilon^0_a}{k_BT}\right) \right\rbrace.
\end{eqnarray}
The equality of these two expressions allows to formulate the Langmuir isotherm such as,
\begin{eqnarray}
\label{eq:O1}
\rho_a & = & \sum^l_{k=1}\rho_{s,k}\ \frac{\frac{1}{k_BT}\ \left(\frac{h^2}{2\pi m k_BT}\right)^{3/2}\ \left[2\sinh\left(\frac{h\nu_{a,k}}{2k_BT}\right)\right]^{-3}\ \exp(\epsilon^0_{a,k}/k_BT)\ P}{1\ +\ \frac{1}{k_BT}\ \left(\frac{h^2}{2\pi m k_BT}\right)^{3/2}\ \left[2\sinh\left(\frac{h\nu_{a,k}}{2k_BT}\right)\right]^{-3}\ \exp(\epsilon^0_{a,k}/k_BT)\ P}.
\end{eqnarray}
The CALF-20 MOF is known to show two CO$_2$ adsorption sites.\cite{nguyen2021,ymagnin2023,drwkeska2024} This is illustrated in the CO$_2$ radial distribution function (RDF) plotted for different $P$, where a peak located at $\sim$ 0.3 nm emerges when $P >$ 0.25 bar, Figure.\ref{fig:figure2}C. The equation \eqref{eq:O1} is thus generalized to a sum of Langmuir's expression, \textit{i.e.} to a multi adsorption sites model,\cite{myers1983} where $k$ denotes the index and $l$ the number of different adsorption sites into the solid host.\\ 
To be of practical interest, the formalism presented above still need to be parametrized by fixing binding energies, saturation densities and vibrational frequencies. To do so, we used the adsorption energy distribution (AED) method,\cite{stanley1993} expressing $\rho_a$ such as,
\begin{eqnarray}
 \label{eq:aed}
\rho_a(P) = \int_{\Gamma} f(\epsilon_a)\ \theta(P,\epsilon_a)\ d\epsilon_a.
\end{eqnarray}
In the AED, the guest density is expressed by summing the fraction of adsorbates coverage $\theta$, weighted by the adsorption energy distribution $f(\epsilon_a)$, spanning over an energy range $\Gamma$. In other words, the AED consists in describing the guest density by summing the probability ($f(\epsilon_a)$) that a molecule implicitly occupies a site in a cage with a corresponding adsorption energy $\epsilon_a$. Thus, from one isotherm, being experimental or from simulations, this method allows in principle to extract parameters like guest-host binding energies, the number of adsorption sites and the saturation density on each site. Such an approach is thus powerful to insight adsorption mechanisms not directly accessible from isotherms, and is of great interest for selecting adsorption models going beyond usual fitting.\cite{gritti2003} The main difficulty in AED consists of determining $f(\epsilon_a)$, which is an unknown function. This can be done thanks to the Expectation-Maximization algorithm.\cite{stanley1993} A section detailing the AED and the way to implement and use it is provided in the supporting information, along with the source code. In the Figure.\ref{fig:figure2}D, we show the energy distribution applied to GCMC isotherm at $T$=293.15K (blue open circles in the Figure.\ref{fig:figure2}A). This is noteworthy that applying the method to GCMC data instead of experimental ones was driven by the prior good agreement found in between both approaches, and most importantly by the large $P$ range spanned by GCMC allowing a robust convergence of the AED, not possible with available experiments, \textit{cf.} dedicated section in the supporting information. The AED distribution shows two peaks (red dashed lines in Figure.\ref{fig:figure2}D), confirming the existence of two adsorption sites in CALF-20 cages, in agreement with the RDF (Figure.\ref{fig:figure2}C). We then determine the corresponding adsorption energies from $f(\epsilon_a)$, namely, $\epsilon_{a,1}$ and $\epsilon_{a,2}$. To correct energies obtained from the AED (corresponding to finite $T$), we express $P$ from \eqref{eq:mua} and \eqref{eq:mug},
\begin{eqnarray}
\label{eq:pressure}
P &=& k_BT\ \frac{\theta}{1-\theta}\ \left(\frac{h^2}{2\pi m k_BT}\right)^{-3/2}\ \left[2\sinh\left(\frac{h\nu_a}{2k_BT}\right)\right]^{3}\ \exp\left(-\frac{\epsilon^0_a}{k_BT}\right).
\end{eqnarray}
Using this latter in Clausius-Clapeyron's law, it comes after few algebra that,
\begin{eqnarray}
\label{eq:bindingE}
\left.-\Delta H\right|_{\rho_a} &=& \epsilon_{a}\ = \ \epsilon^0_{a}\ -\ \frac{k_BT}{2}\ -\ \frac{k_BT^2}{\rho_s - \rho_a}\ \left.\frac{d \rho_s}{dT}\right|_{\rho_a}.
\end{eqnarray}
The last term in \eqref{eq:bindingE} is usually neglected,\cite{hill1986} since the saturation density in CALF-20 is almost independent of $T$, Figure.\ref{fig:figure3}A.\\
Integrating \eqref{eq:aed} with $\theta$=1 then allows to determine the saturation density $\rho_s$ (in the two sites), while integrating each site separately allows to determine the saturation on each site, namely, $\rho_{s,1}$ and $\rho_{s,2}$.\\ 
Vibrational frequencies are approximated by a spring with $\nu_{a,k} \sim \sqrt{\epsilon_{a,k}/ m}/2\pi l_{a,k}$, with $l_{a,k}$ the amplitude of the CO$_2$ vibrations when adsorbed on the solid surface.\cite{hill1986} This amplitude is averaged by molecular dynamics (MD) simulation and found at around $l\sim$ 0.105 nm, SI.Figure.2. This is in line with a common approximation consisting of adjusting the molecular vibrational amplitude to the molecular size $\sigma/2$. From the two binding energies above, we thus determine the two vibrational frequencies, found at the scale of THz, a value expected remembering that the molecular relaxation time is at around the picosecond. As additional benchmark, these two values are shown to be in a reasonable agreement with the CO$_2$ vibration spectrum determined from the Fourier transform of the CO$_2$ velocity autocorrelation function by MD, SI.Figure.3. All parameters discussed above are listed in the Table.\ref{tbl:param_aed}.

\begin{center}
\begin{table}[h]
\caption{\ Parameters of expression \eqref{eq:O1} from the AED method applied to GCMC isotherms. The last column corresponds to parameters directly obtained from GCMC isotherm and enthalpy of adsorption.}
\label{tbl:param_aed}
\begin{tabular}{|c||c|c|}
\hline 
 							& AED 	& GCMC \\ 
\hline 
\hline  
$\rho_s$ (mmol/g) 			& 5.73 			& 5.65\\ 
\hline 
$\rho_{s,1}$ (mmol/g) 		& 2.82 			& 3\\ 
\hline 
$\rho_{s,2}$ (mmol/g) 		& 2.91 			& 2.65\\ 
\hline
$\epsilon^0_{a,1}$ (kJ/mol) & 37.42			& 37.6 \\ 
\hline
$\epsilon^0_{a,2}$ (kJ/mol) & 27.18 		& -\\
\hline
$\nu_{a,1}$ (THz) 			& 1.40 			& -\\
\hline
$\nu_{a,2}$ (THz) 			& 1.19 			& -\\
\hline
\end{tabular} 
\end{table}
\end{center}

\begin{figure}[h!]
\centering
\includegraphics[scale=0.27]{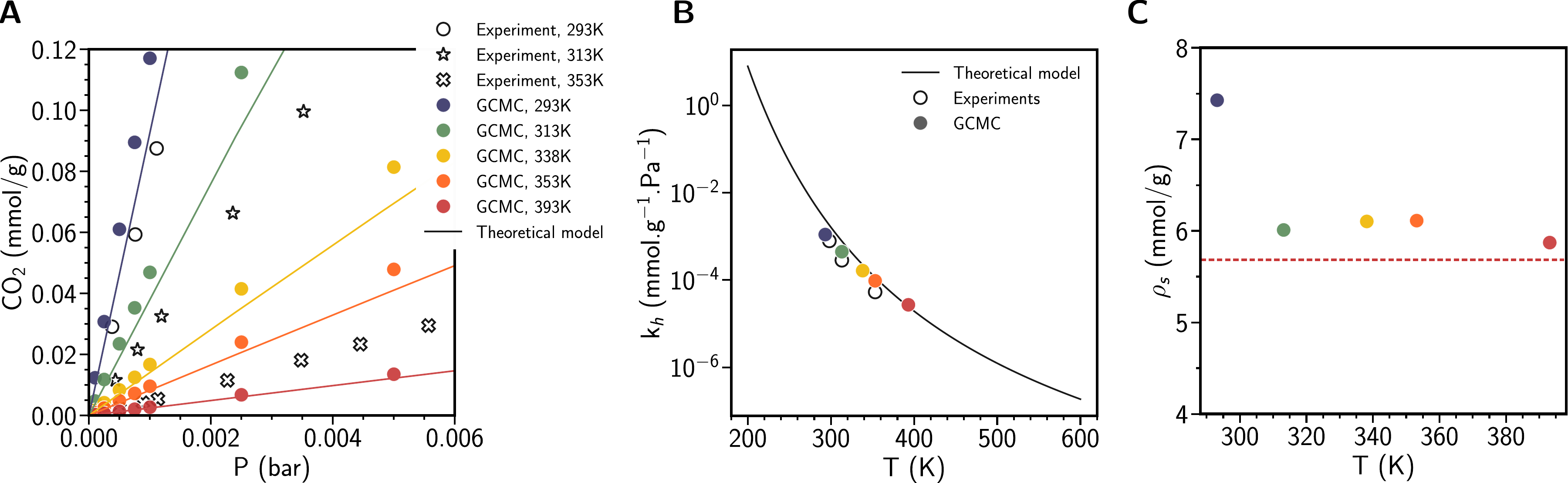}
\caption{A. CO$_2$ isotherms by GCMC (colored circles), experiments (open symbols) and theoretical model (colored lines) in low pressure conditions. GCMC isotherms are shown at different temperatures, $T$=293.15K (blue circles), 313.15K (green circles), 338.15K (yellow circles), 353.15K (orange circles) and $T$=393.15K (red circles). Experimental isotherms are shown at $T$=293.15K (open circles), 313.15K (open stars) and 353.15K (open crosses). B. Log-lin plot of the Henry's constant through different temperatures by GCMC (colored circles), experiments (open symbols) and from the theoretical model (solid line). C. Saturation density through different temperatures by GCMC (red dashed line) and from the theoretical model (colored circles).}
\label{fig:figure4} 
\end{figure}

\noindent The resulting parameters are then used in the adsorption model \eqref{eq:O1} (solid lines) and compared to GCMC simulations (colored circles) at different $T$ on different range of $P$, Figure.\ref{fig:figure3}A,B. As demonstrated, this approach allows to approximate one isotherm at a given $T$ to an other.\\ 
We can then go a step further and predict other thermodynamical parameters of interest such as the Henry's coefficient. At the low pressure regime, the dual Langmuir's isotherm is written as, $\rho_a(P \rightarrow 0)\ =\ \rho_{s,1}\times a_1\times P\ +\ \rho_{s,2}\times a_2\times P= k_H\times P$, and thus $k_H=\rho_{s,1}\times a_1\ +\ \rho_{s,2}\times a_2$, with $a$ the equilibrium constants in \eqref{eq:langmuir_orig}, corresponding in expression \eqref{eq:O1} to,
\begin{eqnarray}
\label{eq:a}
a_k & = & \frac{1}{k_BT}\ \left(\frac{h^2}{2\pi m k_BT}\right)^{3/2}\ \left[2\sinh\left(\frac{h\nu_{a,k}}{2k_BT}\right)\right]^{-3}\ \exp(\epsilon^0_{a,k}/k_BT).
\end{eqnarray}
Coupling these last expressions together, we can express the Henry's coefficient such as,
\begin{eqnarray}
\label{eq:kH}
k_H & = & \sum^2_{k=1}\ \frac{\rho_{s,k}}{k_BT}\ \left(\frac{h^2}{2\pi m k_BT}\right)^{3/2}\ \left[2\sinh\left(\frac{h\nu_{a,k}}{2k_BT}\right)\right]^{-3}\ \exp(\epsilon^0_{a,k}/k_BT),
\end{eqnarray}
where we recognized the Arrhenius expression usually expressed, $k_H=k_\infty \exp(-\Delta H/k_BT)$. Using \eqref{eq:kH} (solid lines) we show a reasonable agreement with GCMC simulations for the different temperatures (colored circles), isotherms at low loading Figure.\ref{fig:figure4}A and Henry's constant, Figure.\ref{fig:figure4}B. It is worth noting that inversely, if $k_H$ is known, $\rho_s$ may be approximated in a straightforward manner from \eqref{eq:kH}. This is shown in the Figure.\ref{fig:figure4}C, where the prediction of the model (colored circles) is close to GCMC densities (red dashed line). We should however notice that such an approach, based on Langmuir's formalism, neglects guests lateral interactions (shown above to induce a decrease in $\Delta H$) does not allow to determine the heat of adsorption.

\clearpage
\section{Kinetics}
\subsection{CO$_2$ diffusion in CALF-20(Zn) by atomistic simulations}
\noindent Before approaching diffusion through theory, the CO$_2$ self-diffusion $D_s$ was investigated by MD simulations with the mean square displacement technique (blue circles $T$=293.15K, red circles $T$=393.15K), Figure.\ref{fig:figure5}A. In it $D_s$ shows a decrease as a function of $P$ for the two $T$ considered. At low loading, the large free pore volume coupled to smaller enthalpy (less negative value) of adsorption along the $P$ range explored (Figure.\ref{fig:figure2}B) allows molecules to hop from one adsorption site to another with a maximum mobility. When the loading increases, the free pore volume is gradually reduced resulting in the decrease of $D_s$. Interestingly, we can note that $D_s$ strongly decreases bellow $P\sim$ 0.25 bar, a situation corresponding to a large occupation of the first adsorption sites (Figure.\ref{fig:figure2}C), then smooth at larger $P$. This is further confirmed by the coupled behaviors of the adsorbates escape time $\tau_e$ (corresponding to the time in between two molecular jumps) and the molecular mean free path $\lambda$ (jump length). These two parameters were averaged from individual molecular displacements by MD simulations and plotted as a function of $P$ in Figure.\ref{fig:figure5}B and its inset. Hence, at low loading, the jump length is $\sim$ 0.6 nm, decreasing to $\sim$ 0.3 nm at larger $P$, both distances corresponding almost to the first and second peaks in the RDF (Figure.\ref{fig:figure2}C). As a result, when $P$ increases, $\tau_e$ is shown to increase, \textit{i.e.} molecules are adsorbed for a longer time between two jumps while their displacement lengths $\lambda$ decrease.

\subsection{CO$_2$ diffusion in CALF-20(Zn) by statistical mechanics}
\noindent In nanopores with sizes of about few molecular diameters, molecules cannot escape the force field induced by solid pore surfaces, that originates a stochastic diffusion where molecules move through successive jumps from one adsorption site to an other into the porous network. The probability of jumping is thus described by an activated process,\cite{beerdsen2004,mo1991,verstreken2024,schlaich2025} where one adsorbed molecule will occasionally have a sufficient energy to leave it and relocates to a neighbor site, or come back to its prior position. Fundamentally, to diffuse, a molecule confined in pores of few molecular diameters may overcome an energy barrier separating two adsorption sites. Such a diffusional mechanism is described by the so-called transition state theory (TST),\cite{camp2016,dubbeldam2005,schlaich2025} where the molecular hoping rate $\nu_e$ is expressed by,
\begin{eqnarray}
\label{eq:tst}
\nu_e &=& \frac{1}{\tau_e}\ =\ \kappa\ \frac{1}{2}\left(\frac{2\ k_BT}{\pi m}\right)^{1/2} \frac{\exp[-\Delta \mathcal{F}(r_b)/k_BT]}{\int_\mathcal{L} \exp[- \Delta \mathcal{F}(r)/k_BT]\ dr}.
\end{eqnarray}
In this one dimensional expression, the square root corresponds to the thermal velocity (at the top of the energy barrier at a position denoted by $r_b$), multiply by the probability that a molecule jumps from a site located at the position $r$ to an other at $r \pm dr$. The jump probability is described by the exponential ratio in \eqref{eq:tst}, a function of the size of the free energy barrier $\Delta \mathcal{F}(r_b)$, normalized by the sum of the exponential of the free energy microstates visited during a jump over a distance $\mathcal{L}$ corresponding to the length span by the barrier. The factor $1/2$ is a normalization factor denoting the equiprobability that the hoping molecule goes in the positive or in the negative direction ($\pm\ dr$) and $\kappa \in$ [0;1] is the Bennett-Chandler correction factor,\cite{chandler1978} accounting for molecules recrossing the energy barrier during a move. Using the Einstein-Smoluchowski relation (ES),\cite{karger2012,schlaich2025} we can write the self-diffusion coefficient such as $D_s=\lambda^2/(2d)\times \nu_e$, where $d$ corresponds to the dimension of the system. Thanks to ES and \eqref{eq:tst}, the self diffusion coefficient can be expressed such as,
\begin{eqnarray}
\label{eq:kin3}
D_s &=& \kappa\ \frac{\lambda^2}{2d}\ \left(\frac{k_BT}{2\pi m}\right)^{1/2} \frac{\exp[\Delta \mathcal{F}(r_b)/k_BT]}{\int_\mathcal{L} \exp[- \Delta \mathcal{F}(r)/k_BT]\ dr}.
\end{eqnarray}
\begin{figure}[h!]
\centering
\includegraphics[scale=0.35]{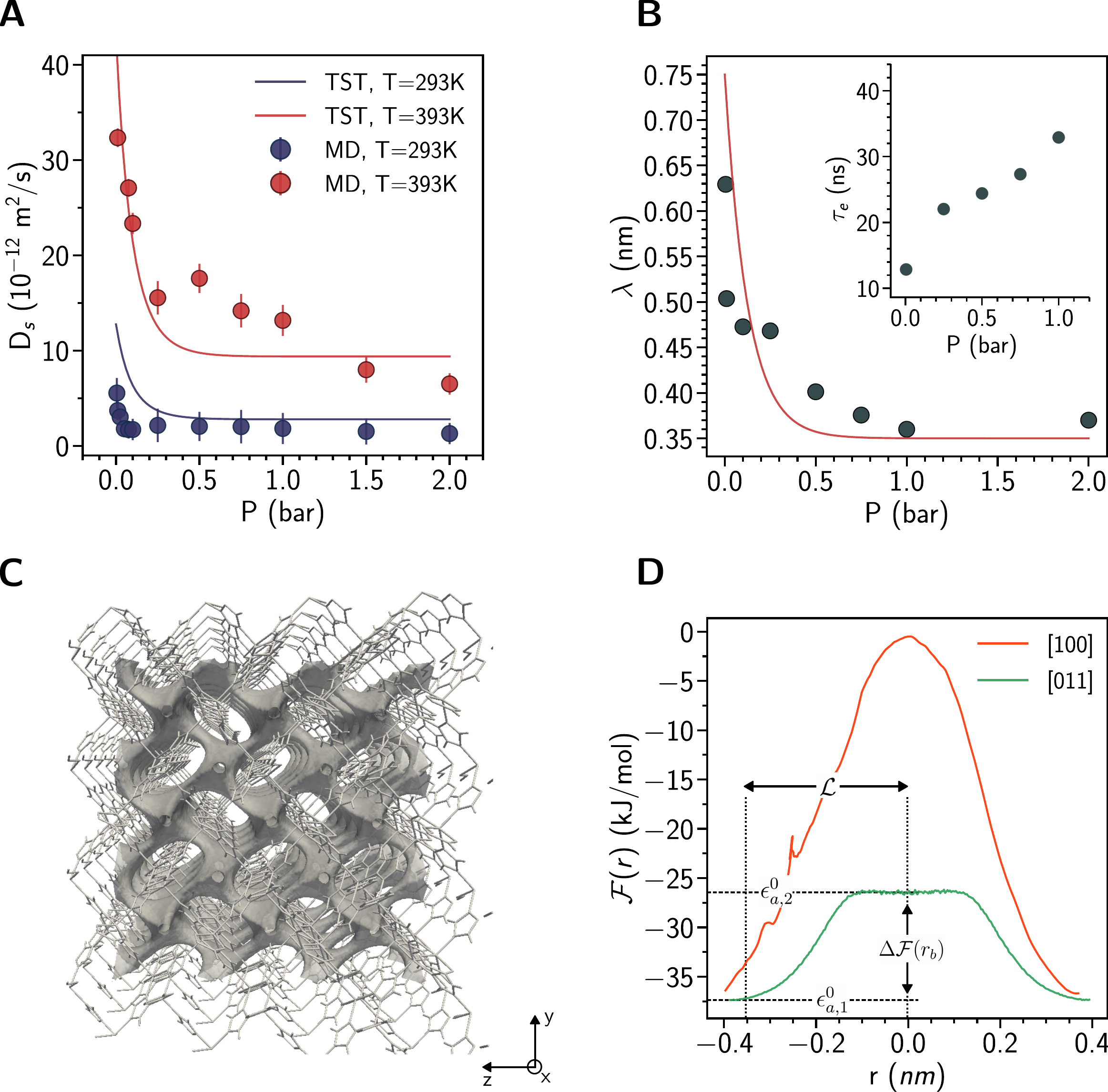}
\caption{A. CO$_2$ self-diffusion coefficient in rigid CALF-20 as a function of $P$ from molecular dynamics at $T$=293.15K (blue circles), $T$=393.15K (red circles) and from the TST expression at $T$=293.15K (blue line) and 393.15K (red line). B. CO$_2$ mean free path as a function of $P$ by molecular dynamics ($T$=293.15K). The inset corresponds to the CO$_2$ escape time as a function of $P$ by molecular dynamics in the same conditions. C. Tree dimensional CALF-20 pore network "has seen" by a CO$_2$ molecule. D. Energy barrier separating two CALF-20 cages along the [100] direction (red line) and on [011] direction (green line).}
\label{fig:figure5}
\end{figure}
In the case of porous crystalline solids like MOF, the mean free path $\lambda$ corresponds almost to the distance that separates neighboring adsorption sites. We then determined $\Delta \mathcal{F}(r)=\mathcal{F}(r)-\mathcal{F}(r_0)$ (with $r_0$ denoting the bottom of the energy barrier) by exploring the potential energy with a probe molecule along the possible diffusion paths in CALF-20 pores, Figure.\ref{fig:figure5}C. Whithin it, the molecule can diffuse through channels connecting CALF-20 cages along the [011] diagonal and/or on the [100] direction. Collecting energies along these paths, we found barriers corresponding to $\Delta \mathcal{F}(r_b)_{y,z}\sim$11 kJ/mol (green line) and $\Delta \mathcal{F}(r_b)_{x}\sim$38 kJ/mol (red line) in Figure.\ref{fig:figure5}C. In the particular case of CALF-20, the two adsorption sites described above are found to be located on the diffusion path in [011] and thus the energy barrier may be linked to thermodynamical parameters such as, $\Delta \mathcal{F}(r_b)_{y,z}\sim\epsilon^0_{a,1}-\epsilon^0_{a,2}$. Note this may not be true for each MOF materials. To predict the self-diffusion beyond infinite dilution, we crudely approximate the effect of finite $P$ from the mean free path as a decreasing exponential varying from $\lambda \sim$ 0.6 nm to 0.35 nm (red line) in Figure.\ref{fig:figure5}B. The expression \eqref{eq:kin3} is then compared to the mean square displacement computed by MD with $\kappa$=1 at $T$=293.15K (blue line) and 393.15K (red line), Figure.\ref{fig:figure5}A. Doing so, we obtained a self-diffusion of about $10^{-12}$ m$^2$/s on the [011] direction and about $10^{-17}$ m$^2$/s in [100]. In this approach several assumptions have been made. We fixed $\kappa$ to the unity for a sack simplicity, neglecting the recrossing effect. We also considered the energy barrier independent of $T$, neglecting the effect of phonons (which may dynamically modify the energy barriers) considering a rigid MOF structure (no atom vibrations). This latter is well known to often notably shift up or down the diffusion coefficient (increasing $D_s$ by one order of magnitude in CALF-20\cite{ymagnin2023}). Although these points may be accounted by carefully determining energy barriers for each considered loading by numerical simulations, this however remains out of the scope of this work.

\section{Conclusion}
\noindent In this work we used both atomistic simulations and statistical mechanics to propose an approach allowing to predict CO$_2$ isotherms and approximate diffusion coefficients as a function of the pressure at different temperatures, from a minimal set of inputs. To do so, Monte Carlo simulations (prior shown to be in good agreement with experiments) and molecular dynamics were first used as numerical experiments to extract adsorption parameters by the adsorption energy distribution technique, being then used to feed and benchmark the dual Langmuir isotherm written from statistical mechanics approach. From the thermodynamics standpoint, this isotherm was shown to depend, in addition to the binding energy with the solid surface, on the molecular vibrations (on its adsorption site) originated by the thermal excitations. This formalism and the associated methodology was shown to enable the extrapolation of isotherm, Henry's constant and saturation densities from one $T$ ranging from 293.15 to 393.15K. In a second time, we focused on CO$_2$ diffusion based on the transition state theory feed by some parameters prior determined by thermodynamics. We hope that such a work may be appealing for porous materials engineering community and help in shedding light into intrinsic mechanisms driving molecular diffusion in strongly confined environments.

\section{Methods}
\noindent Adsorption and diffusion simulations were performed by GCMC from RASPA\cite{dubbeldam2016} and MD from LAMMPS\cite{thompson2022}, respectively in CALF-20 structure provided by Shimizu \emph{et al.}\cite{lin2021}. In simulations we considered CO$_2$ interacting into CALF-20 MOF structure\cite{lin2021} of 6$\times$11.8$\times$3.9 nm edges at $T$=293.15K. The framework was considered as rigid (\textit{i.e} each atom are frozen) for both GCMC and MD calculations. Atomistic parameters for all CALF-20 atoms were taken from the DREIDING force field (provided in the supporting information),\cite{mayo1990} with point charges previously computed from the REPEAT method\cite{campana2009} as proposed by Shimizu \textit{et al.}\cite{lin2021} The non bonded guests-host interactions were ensured by a Lennard-Jones potential for CO$_2$ with the parameters proposed by Garci\'a-S\'anchez \textit{et al.}\cite{garcia2009} (provided in the supporting information), with Lorentz-Berthelot combining rules for cross interactions, with a cutoff fixed at 1.2 nm. In addition, long range electrostatic interactions were ensured by the PPPM algorithm,\cite{hockney2021} with an accuracy of $10^{-6}$. Such a set of parameters was already been shown in CALF-20 MOF to allow a good agreement in between simulations and experiments for CO$_2$ isotherms and enthalpy of adsorption.\cite{ymagnin2023,gopalsamy2024}. Each pressure presented in this work were corrected by the Peng-Robinson equation of state\cite{peng1976} in GCMC simulations, the system was equilibrated during 10$^6$ Monte Carlo cycles, the first $10^5$ cycles used for equilibration and next cycles for averaging equilibrium parameters (guests loading and enthalpy of adsorption). Diffusion was then performed from MD simulations with the LAMMPS package\cite{thompson2022}. Diffusion coefficients were determined from the mean square displacement\cite{karger2012} in the canonical ensemble along 100 of nanoseconds runs with a time step fixed at 0.5 fs. Each diffusion plot proposed were reproduced 3 times from the different initial configurations randomly selected from equilibrated microstates by prior GCMC runs. In all MD simulations, a Nose-Hoover thermostat were used to keep constant temperature during simulations. In order to plot the vibration spectrum, MD simulations were first used to compute the velocity auto-correlation (VACF) of CO$_2$ centers of mass, $1/N \sum_{i=1}^N \textbf{v}_i(t=0).\textbf{v}_i(t)$. To do so, we record the C and O velocities each 5 fs in a 2 ns MD runs and the VACF was averaged 100 times. From these latter we then computed the VACF Fourier transform corresponding to the vibrational spectrum.

\section*{Acknowledgments}
This work was supported by TotalEnergies S.E. through OneTech, Sustainability, CCUS R\&D program. On the behalf of all co-authors, Y. Magnin thanks the company Decarbontek LLC for kindly sharing CALF-20 batch sample used for experiments presented in this work, S. Frambati and B. Coasne for useful discussions and TotalEnergies High Performance Computing Center for the computational time provided on its super calculator PANGEA, and more particularly N. Gohaud, E. Bergounioux and D. Klahr for there key help in solving technical issues in running calculations on the novel cluster PANGEA IV.

\clearpage
\section*{Supporting Information: Fundamental of CO$_2$ Adsorption and Diffusion in Sub-nanoporous Materials: Application to CALF-20}
\begin{figure}[!ht]
\centering
\includegraphics[scale=0.5]{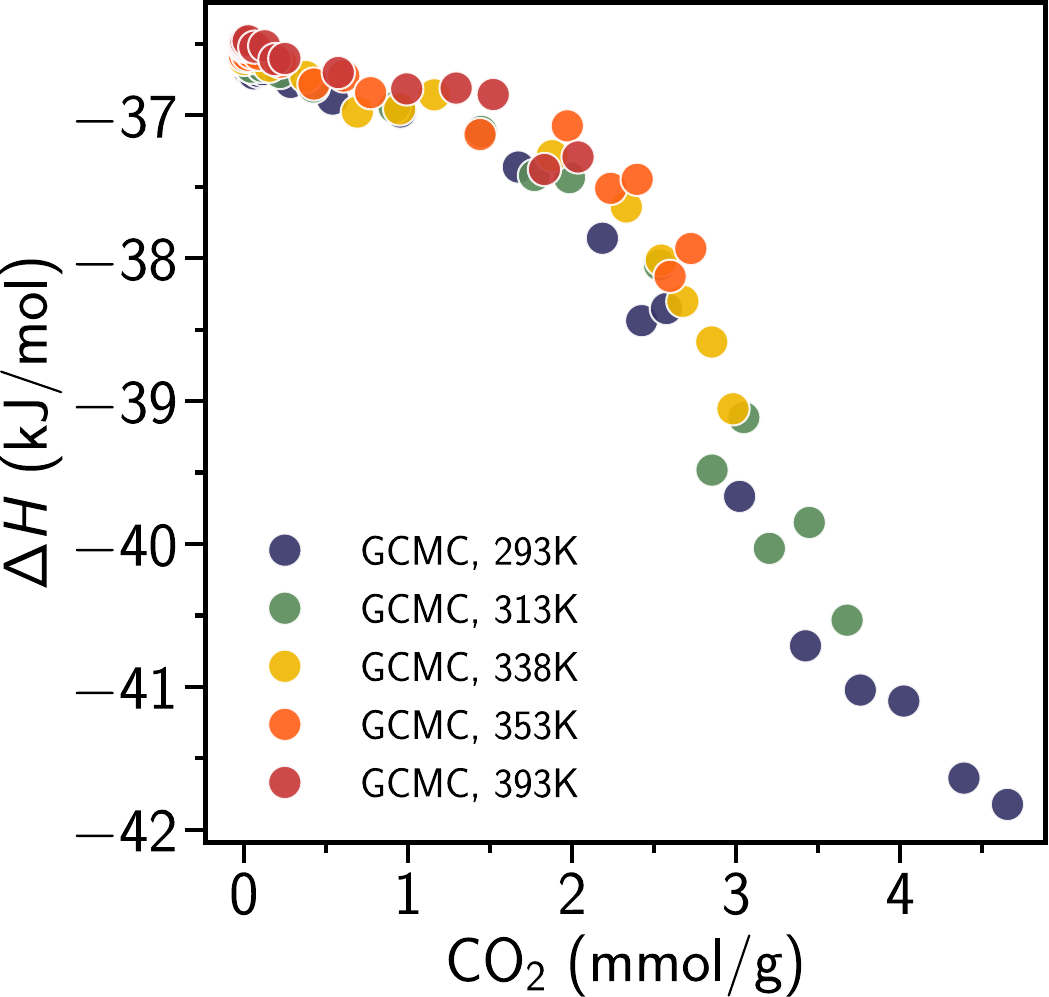}
\caption{Enthalpy of adsorption as a function of the CO$_2$ loading by GCMC simulations at $T$=293.15K (blue circles), 313.15K (green circles), 338.15K (yellow circles), 353.15K (orange circles) and $T$=393.15K (red circles).}
\label{fig:SI_figure1}
\end{figure}

\begin{figure}[!ht]
\centering
\includegraphics[scale=0.5]{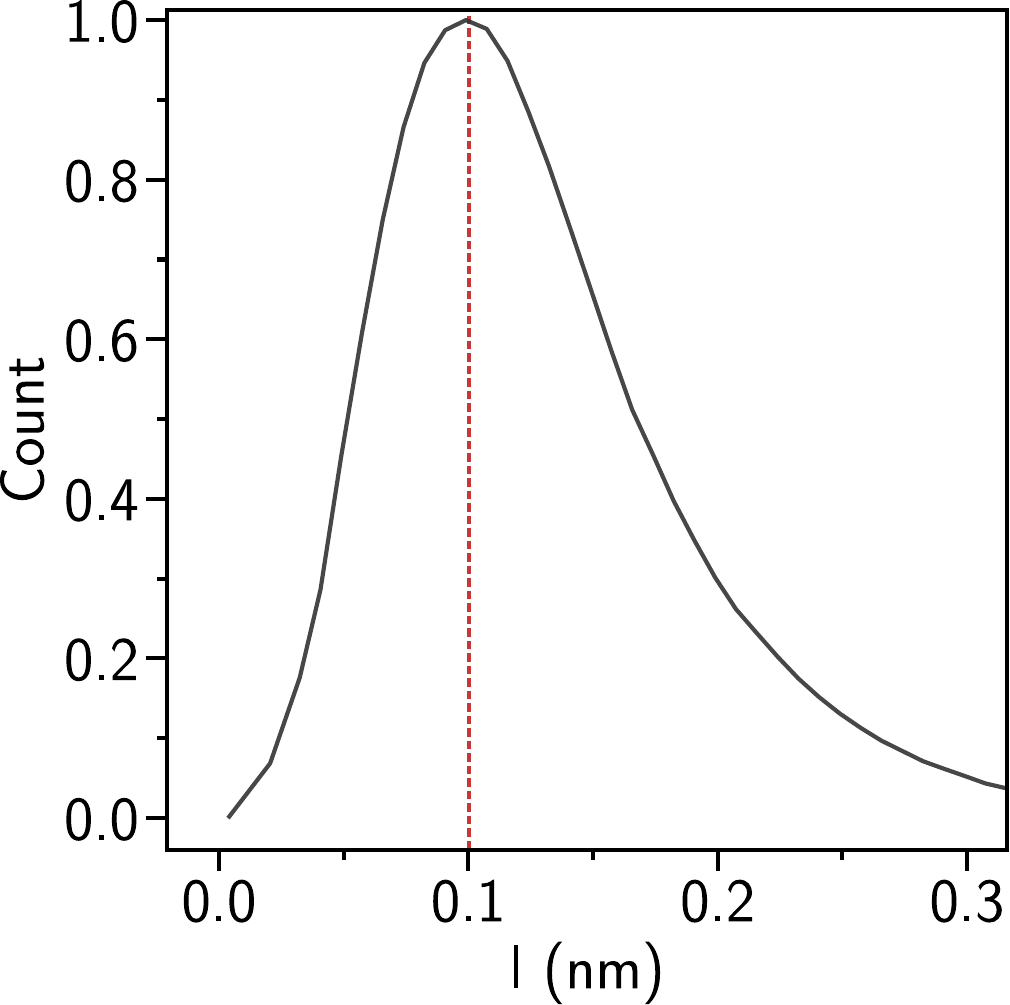}
\caption{Distribution of the CO$_2$ amplitude vibrations in CALF-20 MOF at $T$=293.15K and $P$=0.01 bar.}
\label{fig:SI_figure2}
\end{figure}

\begin{figure}[!ht]
\centering
\includegraphics[scale=0.5]{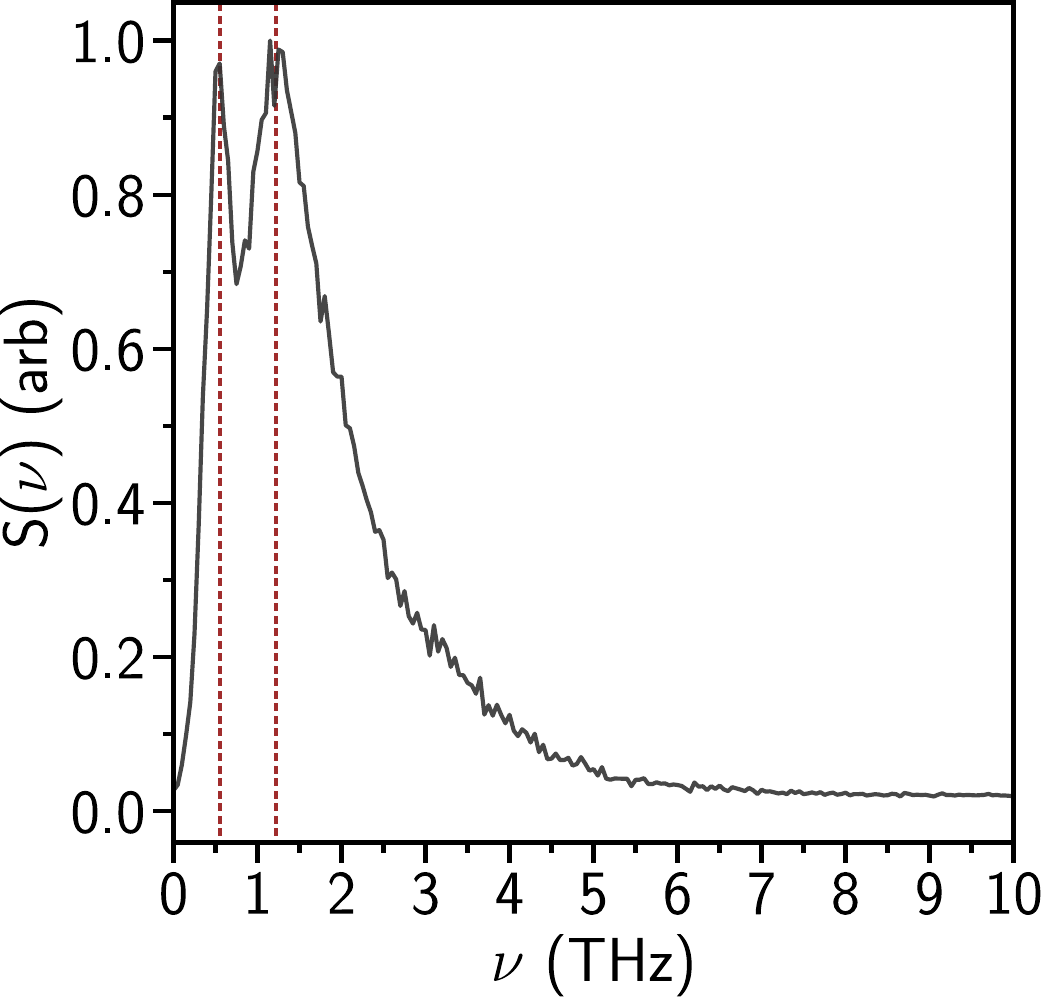}
\caption{CO$_2$ vibration spectrum at $T$=293.15K and $P$=0.01 bar.}
\label{fig:SI_figure3}
\end{figure}

\clearpage

\section{Adsorption energy distribution (AED)}
\subsection{The Expectation-Maximation algorithm for AED determination}
The numerical estimation of the adsorption energy distribution is performed on the basis of the Expectation-Maximation (EM) method adapted for parameter estimation, as presented by Stanley and Guiochon.\cite{stanley_guiochon1993} If we initially assume that $f(\epsilon_a)$ is the probability of occupation of an adsorption site of energy $\epsilon_a$, the density of an adsorbed solute at bulk pressure $P$ may be approximated as
\begin{equation}
    \rho_a(p)=\sum_{\epsilon_{a_{min}}}^{\epsilon_{a_{max}}}f(\epsilon_a)\theta(p,\epsilon_a)\Delta\epsilon_a,
    \label{eqs1}
\end{equation}
where $\epsilon_{a_{min}}$ and $\epsilon_{a_{max}}$ are the lower and the higher energy boundary of the adsorption sites, respectively, and $\theta$ is the fraction of surface coverage of the homogeneous or local model of adsorption (\textit{e.g.}, Langmuir model). The EM method seeks the distribution function $f(\epsilon_{a})$ that minimizes the deviations between the experimental (or simulated) guest density and the one predicted from \eqref{eqs1}. To do so, the energy space is discretized in $M$ grid points separated by the energy interval $\Delta\epsilon_a$, with
\begin{equation}
    \Delta\epsilon_a=\frac{\epsilon_{a_{max}}-\epsilon_{a_{min}}}{M}.
    \label{eqs2}
\end{equation}
For each energy level $\epsilon_{a}$, $f(\epsilon_{a})$ is iteratively updated as
\begin{equation}
    f^{k+1}(\epsilon_{a})=f^{k}(\epsilon_{a})\frac{\sum\limits_{p_j=p_{\text{min}}}^{p_{\text{max}}}\theta(p_{j},\epsilon_{a})\Delta \epsilon_{a}\frac{\rho_a^{\text{exp}}(p_j)}{\rho_a^{\text{calc}}(p_j)}}{\sum\limits_{p_j=p_{\text{min}}}^{p_{\text{max}}}\theta(p_{j},\epsilon_{a})\Delta \epsilon_{a}},
    \label{eqs3}
\end{equation}
whereby $f^{k}(\epsilon_a)$ is corrected, at every iteration $k$, by the ratio between experimental and calculated guest densities, averaged in the whole pressure range covered by the adsorption dataset. To keep the equation dimensionally consistent, we have included the factor $\Delta \epsilon_{a}$ in the denominator. The iterative process is maintained until the square of deviations between experimental and calculated guest densities becomes less than or equal to the desired tolerance ($tol$):
\begin{equation}
    \sum\limits_{p_j=p_{min}}^{p_{max}}\left(\rho_a^{\text{calc}}(p_j)-\rho_a^{\text{exp}}(p_j)\right)^{2}\leq tol  
    \label{eqs4}    
\end{equation}
To avoid biased estimations, our choice of initial guess for $f(\epsilon_{a})$ is adapted from Stanley and Guiochon,\cite{stanley_guiochon1993} in which the maximum loading provided in the experimental dataset is evenly distributed among the $M$ grid points of the energy space, as follows:  
\begin{equation}
    f(\epsilon_{a}) = \rho_{a}(p_{\text{max}})/M/\Delta \epsilon _{a}  
    \label{eqs5}    
\end{equation}
The saturation density ($\rho_s$) can be computed by setting $\theta\ = 1$ (full surface coverage) in equations \eqref{eqs1}, leading to:
\begin{equation}
    \rho_s=\int\limits_{\epsilon_{a_{\text{min}}}}^{\epsilon_{a_{\text{max}}}}f(\epsilon_a)\ \text{d}\epsilon_a,
    \label{eqs6}
\end{equation}
where the integral must be numerically evaluated.\\
In this work, isotherms are assumed to be of type I, so that the Langmuir model is used as the local adsorption model,
\begin{equation}
    \theta(P,\epsilon_{a})=\frac{a P}{1+a P}.
\end{equation}
The equilibrium constant $a$ is determined from the vapor pressure of the adsorbate at experimental temperature ($P_{v}$), and from its molar enthalpy of vaporization ($\Lambda$):
\begin{equation}
    a^{-1}=P_{v}\ \exp(\Lambda/k_BT).
\end{equation}
For CO$_2$, both the enthalpy of vaporization and the vapor pressure are taken from the NIST database.\cite{nist} We have neglected the temperature dependence of the enthalpy of vaporization, setting $\Lambda$ = 16.7 kJ/mol - value specified at 288 K -  in all calculations.\cite{nist}\\
In the remainder of this material, we present the results of the calculations performed with the EM method. Initially, the AED is calculated from experimental datasets limited to 1 bar. We show that the range of data used as input has a great impact in the accuracy of results. In the final section, the AED is calculated from a dataset generated from molecular simulations performed in the grand-canonical ensemble (GCMC), and further validated against experimental data. A code written in FORTRAN 90 is made available together with this material under the MIT License on GitHub \href{url}{https://github.com/LESC-Unicamp/Supplementary-materials/tree/main/Fundamental\%20of\%20CO2\%20Adsorption\%20and\%20Diffusion\%20in\%20Sub-nanoporous\%20Materials\%3A\%20Application\%20to\%20CALF-20}. It reads the input dataset and main parameters from the 'input.dat' file, and prints the energy distribution function and the estimated isotherm in the file 'energydistribution.dat'. Compiling and running instructions are detailed in the file 'tutorial.dat'.

\subsection{AED from experimental datasets}
\begin{figure}[h!]
\centering
\includegraphics[scale=0.6]{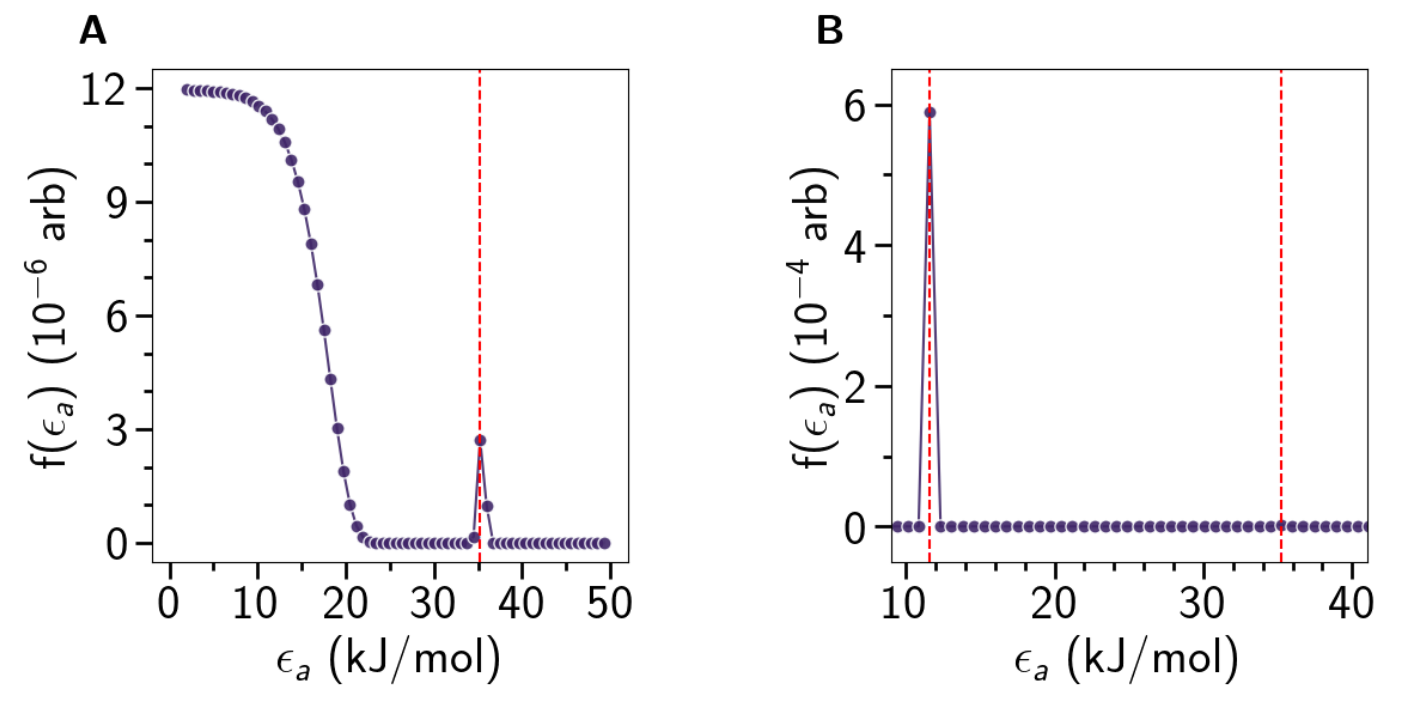}
\caption{Energy distributions estimated from the experimental dataset at 293.15 K ($tol = 3.591\cdot10^{-4}$).  A. Constant initial guess. B. Gaussian distribution as initial guess ($\mu$ = 26 kJ, $\sigma$ = 0.4 kJ).}
\label{fig_aed1} 
\end{figure}

\begin{figure}[h!]
\centering
\includegraphics[scale=0.55]{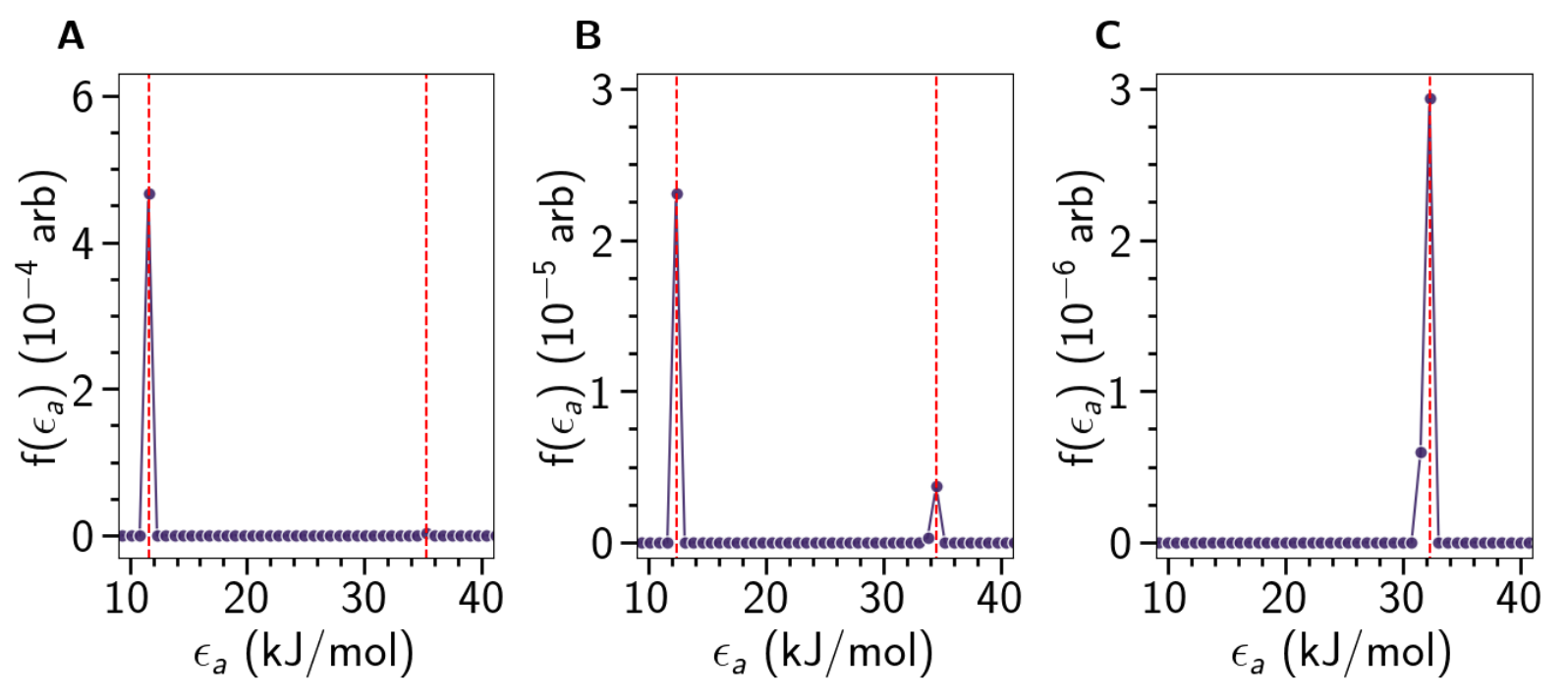}
\caption{Energy distributions estimated from experimental datasets. A single Gaussian distribution ($\mu$ = 26 kJ, $\sigma$ = 0.4 kJ) is used as initial guess, in all cases. A. 298.15 K, $tol = 3.430\cdot10^{-4}$. B. 313.15 K, $tol = 3.313\cdot10^{-4}$. C. 353.15 K, $tol = 3.117\cdot10^{-4}$.}
\label{fig_aed2} 
\end{figure}

The first dataset comprises 65 pressure points experimentally determined at 293.15 K, limited to 1 bar. The initial distribution is computed with equation \eqref{eqs5}, assuming an energy range varying from 2 kJ to 50 kJ. The estimated AED is shown in Figure \ref{fig_aed1}A. A second test is performed by setting the initial distribution function as a Gaussian distribution ($\mu$ = 26 kJ, $\sigma$ = 0.4 kJ). The results are illustrated in Figure \ref{fig_aed1}B. 
In Figure \ref{fig_aed1}A, a peak close to 35.2 kJ/mol can be identified, but no second peak can be detected up to 20 kJ. In fact, the distribution estimated in such region is rather unreasonable, indicating a continuum of adsorption sites. This distribution clearly contradicts the radial distribution function estimated by simulations, which indicates the existence of two adsorption sites. When a Gaussian distribution is set as initial guess, a second peak emerges at 11.6 kJ. From the estimated AED, the saturation density calculated with a Gaussian quadrature is close to 437.5 mmol/g - a value that overestimates the saturation density expected from GCMC simulations ($\sim$5.65 mmol/g). The results obtained from both tests indicate that the estimated AEDs are not accurate enough. The scenario illustrated here matches the description presented by Stanley and Guiochon\cite{stanley_guiochon1993} for the case when input data are missing at the low-energy or, equivalently, at the high-pressure range. In such a case, not only spurious peaks may appear, but also the entire distribution may be compromised.\cite{stanley_guiochon1993} Thus, it is crucial to use a wide range of input data in order to avoid artifacts in the following analyzes.\\
Similar scenarios are found at 298.15 K and 313.15 K: constant initial guesses \eqref{eqs5} result in AEDs similar to the one of Figure \ref{fig_aed1}A, while a Gaussian distribution gives an AED close to that illustrated in Figure \ref{fig_aed1}B. In the latter case, the estimated AEDs are illustrated in Figures \ref{fig_aed2}A and \ref{fig_aed2}B, respectively. At 353.15 K, both constant and Gaussian distributions converge to the same AED, shown in Figure \ref{fig_aed2}C. A single peak can be identified, whose corresponding energy level is $\sim$32.3 kJ, close to the energy levels identified at other temperatures 35.2 kJ, at 293.15 K and 298.15 K, and 34.5 kJ, at 313.15 kJ. Such a result indicates that the existence of an adsorption site with an energy level around 35 kJ is plausible, making sense when compared to the enthalpy of adsorption at low loading reported in the literature and in the Figure.2B in the manuscript. However, the estimated saturation density is limited to the first adsorption site in CALF-20 (2.71 mmol/g), possibly due to the lack of input data at high pressure.        

\subsection{AED from GCMC dataset}
\begin{figure}[h!]
\centering
\includegraphics[scale=0.6]{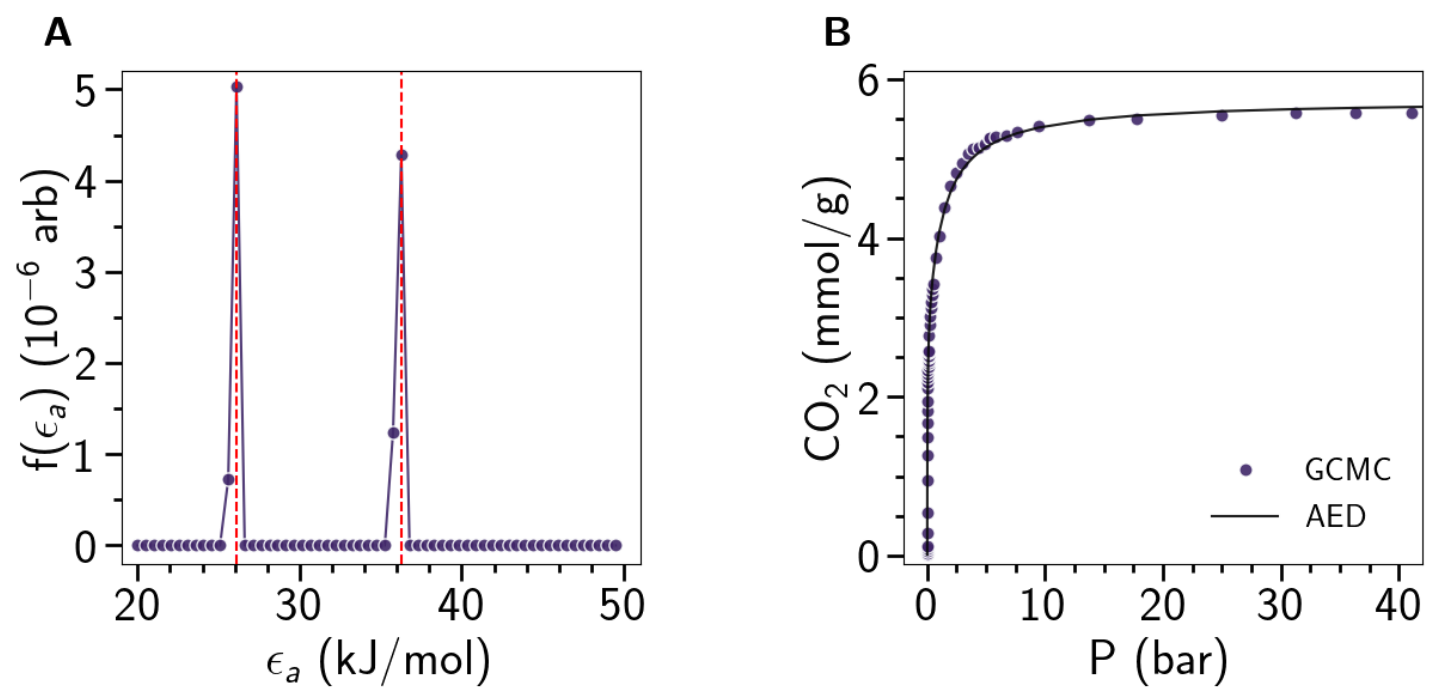}
\caption{A. Energy distribution estimated from the GCMC dataset, at 293.15 K ($tol = 3.591\cdot10^{-4}$, constant initial guess).  B. Adsorption isotherms for CO$_2$: GCMC dataset and the fitting from AED calculations, up to 40 bar.}
\label{fig_aed3} 
\end{figure}

\begin{figure}[h!]
\centering
\includegraphics[scale=0.6]{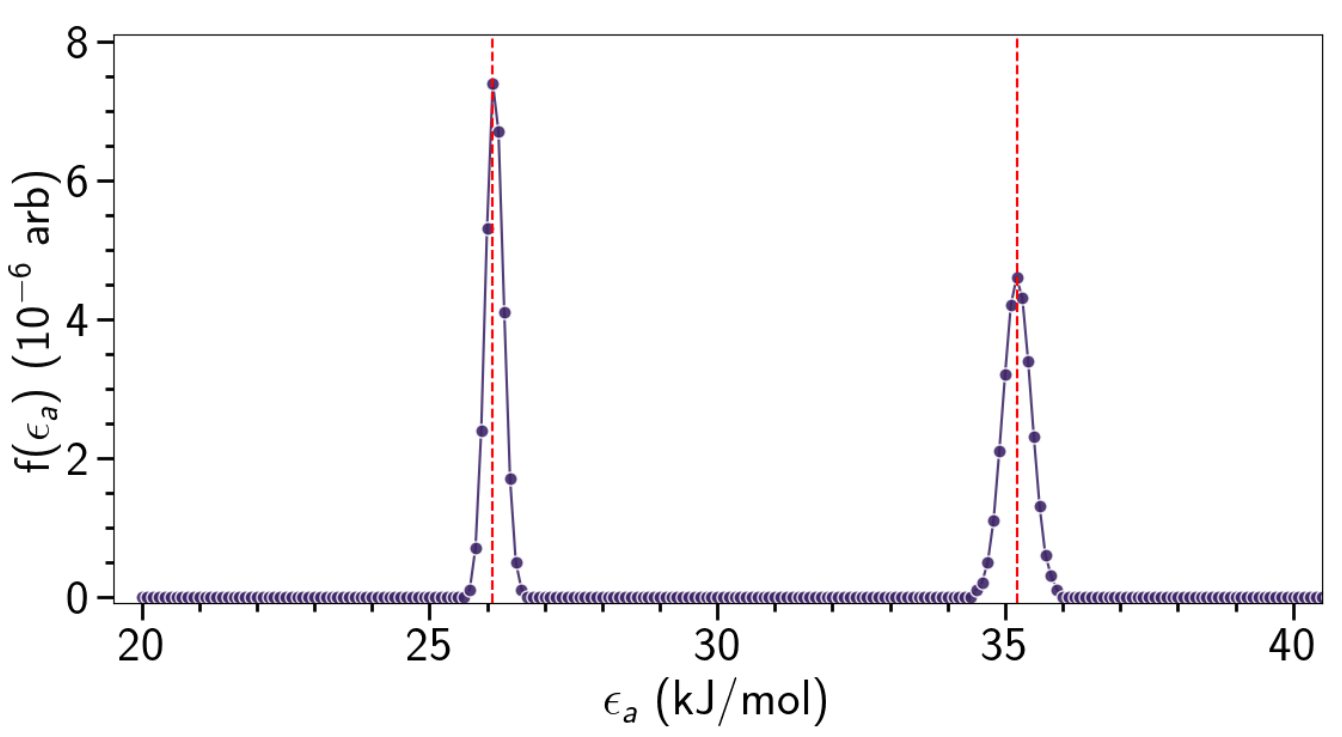}
\caption{Energy distribution estimated from a reverse test: the AED parameters fitted from the GCMC dataset (Figure \ref{fig_aed3}A) are employed in the theoretical model, which is then used to generate a larger dataset, to which AED is further applied ($tol = 4.56\cdot10^{-6}$, constant initial guess.)}
\label{fig_aed4} 
\end{figure}

The generation of data in the high-pressure range of adsorption isotherms by means of experiments can be rather impractical. To circumvent this, we have considered the use of molecular simulations performed in the grand-canonical ensemble (GCMC). An adsorption data set consisting of 59 pressure points at 293.15 K, from $\cdot10^{-4}$ to 2.94$\cdot10^{3}$ bar, is obtained from the simulations and used as input in the EM algorithm. The estimated AED and the corresponding adsorption isotherm are presented in Figures \ref{fig_aed3}A and \ref{fig_aed3}B, respectively.
Two peaks can be identified in Figure \ref{fig_aed3}A: one with $\epsilon_{a}$ = 26.1 kJ/mol (peak 2), and the other with $\epsilon_{a}$ = 36.3 kJ/mol (peak 1). The saturation density calculated from the AED reads as 5.73 mmol/g. By integrating each peak separately, one can determine the saturation densities corresponding to each adsorption site: $\rho_{s,1}$ = 2.82 mmol/g, and $\rho_{s,2}$ = 2.91 mmol/g. Whether starting from a constant initial guess (equation \eqref{eqs5}) or a Gaussian distribution, the final AED remains exactly the same. In order to generate a plot with better resolution, the energy range considered varies from 20 kJ to 50 kJ. Different energy boundaries in the interval 0 kJ/mol to 70 kJ/mol have been also tested, but the final AED didn't change from the one shown in Figure \ref{fig_aed3}A, confirming the existence of two adsorption sites.\\ 
Once the theoretical model proposed in this work was parameterized on the basis of the parameters generated from the AED, an additional test was proposed to attest to the reliability of the method, in which 300 pressure points up to $2.94\cdot10^{3}$ bar were generated from the parameterized model, and reversely served as input to the AED estimation. The converged distribution is shown in Figure \ref{fig_aed4}. Only the energy of the low-pressure adsorption site ($\epsilon_{a,1} = 35.2$ kJ/mol) diverges slightly from the one estimated using the original GCMC dataset, possibly due to the approximations underlying equation (9) of the main text, \textit{i.e.}, the estimation of the guest-host binding energy at 0 K from the one estimated from AED, or even as a result of the estimation of vibrational frequencies in the model, which are not a direct part of AED. The difference in the amount of grid points may also contribute to the small differences observed between Figures \ref{fig_aed3}A and \ref{fig_aed4}.\\
 
\section{Adsorption Experiments}
AUTOSORB iQ from Quantachrome \textsuperscript{\textregistered} was used to perform equilibrium measurements on CALF20 in powdered form. The AUTOSORB is a volumetric or manometric adsorption device which is capable of measuring adsorbed gaseous species at fairly low partial pressures (up to $10^{-7}$ bar), well adapted for Henry's constant measurements presented in this work. This method is based on the measurement of volumes and pressures and the application of the ideal gas equation for the surrounding bulk phase. The adsorbed amount is determined based on the difference between the total amount of gas injected in the sample cell and the amount of gas in a dead space. The sample was pre-treated/activated at 150 degree Celcius under vacuum for 12 hours before running measurements. The dry weight of the CALF20 sample used in CO$_2$ isotherm presented in this work is 0.0979 g (Figure.2A). The sample cell was then immersed in oil bath to regulate the analysis temperature. Before starting an analysis, the automatic procedure includes some necessary preliminary steps listed bellow:
\begin{itemize}
 \item Leak test with Helium.
 \item Volume measurement of the sample cell by Helium at ambient and regulated temperature.
 \item The helium is then removed by vacuum and the dosing system injects the measured gas (CO$_2$ in this case) to fix $P/P_0$ in the sample cell. 
 \item  After injecting the gas in the sample cell, the system is set to the equilibrium. Doing so, an amount of gas will be adsorbed by the sample resulting a pressure decrease in the cell. This loss was calculated to the volume/moles of adsorbed species. Once pressure equilibrium is reached with the desired tolerance (pressure change along with time), a new dose is injected. This step is repeated for different partial pressures ranging from $10^{-5}$ to 1 ($p_{max}$=1 atm).
\end{itemize}
Sixty five equilibrium $P/P_0$ ($P_0$ being 1 atm) were programmed to get a detailed CO$_2$ isotherm in CALF20.

\clearpage
\section{Simulation parameters}
\begin{center}
\begin{table}[h]
\caption{\ CALF-20 Dreiding parameters.\cite{mayo1990} The REPEAT charges were taken from the supplementary materials (Table S3) in Shimizu and co-worker publication.\cite{lin2021}}
\label{tbl:parammof}
\begin{tabular}{|c||c|c|c|c|c|}
\hline 
Atoms  					& Zn & N & O & C & H \\
\hline 
\hline 
$\sigma$ (\AA)   		& 4.0447 & 3.2626 & 3.0331  & 3.473 & 2.8464\\ 
\hline 
$\epsilon$ (kcal/mol)	& 0.055   & 0.0774  & 0.0957    & 0.0951  & 0.0152\\ 
\hline
\end{tabular}
\end{table}
\end{center}

\begin{center}
\begin{table}[h]
\caption{\ Lennard-Jones parameters for CO$_2$ adsorbates from Garci\'a-S\'anchez \textit{et al.}\cite{garcia2009}}
\label{tbl:paramguests}
\begin{tabular}{|c||c|c|}
\hline 
Atoms & C$_{CO_2}$ & O$_{CO_2}$\\
\hline 
\hline 
$\sigma$ (\AA)       & 2.745   & 3.017 \\ 
\hline 
$\epsilon$ (kcal/mol)& 0.0595 & 0.1702\\ 
\hline 
q	                 & 0.6512  & -0.3256  \\
\hline 
\end{tabular}
\end{table}
\end{center}

\clearpage
\bibliographystyle{unsrt}

\begin{thebibliography}{10}

\bibitem{koutsonikolas2016}
D~Koutsonikolas, G~Pantoleontos, M~Mavroudi, S~Kaldis, A~Pagana, ES~Kikkinides,
  and D~Konstantinidis.
\newblock Pilot tests of co 2 capture in brick production industry using
  gas--liquid contact membranes.
\newblock {\em International Journal of Energy and Environmental Engineering},
  7:61--68, 2016.

\bibitem{kearns2021}
David Kearns, Hary Liu, and Chris Consoli.
\newblock Technology readiness and costs of ccs.
\newblock {\em Global CCS institute}, 3, 2021.

\bibitem{orlov2022}
Alexey~A Orlov, Alain Valtz, Christophe Coquelet, Xavier Rozanska, Erich
  Wimmer, Gilles Marcou, Dragos Horvath, B{\'e}n{\'e}dicte Poulain, Alexandre
  Varnek, and Fr{\'e}d{\'e}rick De~Meyer.
\newblock Computational screening methodology identifies effective solvents for
  co2 capture.
\newblock {\em Communications Chemistry}, 5(1):37, 2022.

\bibitem{gao2023}
Huan Gao, Xinke Wang, Kang Wu, Yarong Zheng, Qize Wang, Wei Shi, and Meng He.
\newblock A review of building carbon emission accounting and prediction
  models.
\newblock {\em Buildings}, 13(7):1617, 2023.

\bibitem{perez2022}
Eduardo P{\'e}rez-Botella, Susana Valencia, and Fernando Rey.
\newblock Zeolites in adsorption processes: State of the art and future
  prospects.
\newblock {\em Chemical reviews}, 122(24):17647--17695, 2022.

\bibitem{sircar1996}
S~Sircar, TC~Golden, and MB~Rao.
\newblock Activated carbon for gas separation and storage.
\newblock {\em Carbon}, 34(1):1--12, 1996.

\bibitem{reid1998}
CR~Reid, IP~O'koy, and KM~Thomas.
\newblock Adsorption of gases on carbon molecular sieves used for air
  separation. spherical adsorptives as probes for kinetic selectivity.
\newblock {\em Langmuir}, 14(9):2415--2425, 1998.

\bibitem{li2009}
Jian-Rong Li, Ryan~J Kuppler, and Hong-Cai Zhou.
\newblock Selective gas adsorption and separation in metal--organic frameworks.
\newblock {\em Chemical Society Reviews}, 38(5):1477--1504, 2009.

\bibitem{wang2017}
Yuxiang Wang and Dan Zhao.
\newblock Beyond equilibrium: metal--organic frameworks for molecular sieving
  and kinetic gas separation.
\newblock {\em Crystal Growth \& Design}, 17(5):2291--2308, 2017.

\bibitem{dutcher2015}
Bryce Dutcher, Maohong Fan, and Armistead~G Russell.
\newblock Amine-based co2 capture technology development from the beginning of
  2013-a review.
\newblock {\em ACS applied materials \& interfaces}, 7(4):2137--2148, 2015.

\bibitem{dawson2012}
Robert Dawson, Andrew~I Cooper, and Dave~J Adams.
\newblock Nanoporous organic polymer networks.
\newblock {\em Progress in Polymer Science}, 37(4):530--563, 2012.

\bibitem{zeng2016}
Yongfei Zeng, Ruqiang Zou, and Yanli Zhao.
\newblock Covalent organic frameworks for co2 capture.
\newblock {\em Advanced Materials}, 28(15):2855--2873, 2016.

\bibitem{kumar2015}
Amrit Kumar, David~G Madden, Matteo Lusi, Kai-Jie Chen, Emma~A Daniels, Teresa
  Curtin, John~J Perry~IV, and Michael~J Zaworotko.
\newblock Direct air capture of co2 by physisorbent materials.
\newblock {\em Angewandte Chemie International Edition}, 54(48):14372--14377,
  2015.

\bibitem{nguyen2024_c1}
Tai~TT Nguyen, Bhubesh~Murugappan Balasubramaniam, Nicholas Fylstra, Racheal~PS
  Huynh, George~KH Shimizu, and Arvind Rajendran.
\newblock Competitive co2/h2o adsorption on calf-20.
\newblock {\em Industrial \& Engineering Chemistry Research}, 63(7):3265--3281,
  2024.

\bibitem{rajendran2024_2}
Arvind Rajendran, George~KH Shimizu, and Tom~K Woo.
\newblock The challenge of water competition in physical adsorption of co2 by
  porous solids for carbon capture applications--a short perspective.
\newblock {\em Advanced Materials}, 36(12):2301730, 2024.

\bibitem{polat2024}
H~Mert Polat, Felipe~M Coelho, Thijs~JH Vlugt, Luis~Fernando Mercier~Franco,
  Ioannis~N Tsimpanogiannis, and Othonas~A Moultos.
\newblock Diffusivity of co2 in h2o: A review of experimental studies and
  molecular simulations in the bulk and in confinement.
\newblock {\em Journal of Chemical \& Engineering Data}, 69(10):3296--3329,
  2024.

\bibitem{chanut2017}
Nicolas Chanut, Sandrine Bourrelly, Bogdan Kuchta, Christian Serre, Jong-San
  Chang, Paul~A Wright, and Philip~L Llewellyn.
\newblock Screening the effect of water vapour on gas adsorption performance:
  application to co2 capture from flue gas in metal--organic frameworks.
\newblock {\em ChemSusChem}, 10(7):1543--1553, 2017.

\bibitem{kolle2021}
Joel~M Kolle, Mohammadreza Fayaz, and Abdelhamid Sayari.
\newblock Understanding the effect of water on co2 adsorption.
\newblock {\em Chemical Reviews}, 121(13):7280--7345, 2021.

\bibitem{subramanian2019}
Vishal Subramanian~Balashankar and Arvind Rajendran.
\newblock Process optimization-based screening of zeolites for post-combustion
  co2 capture by vacuum swing adsorption.
\newblock {\em ACS Sustainable Chemistry \& Engineering}, 7(21):17747--17755,
  2019.

\bibitem{subraveti2019}
Sai~Gokul Subraveti, Kasturi~Nagesh Pai, Ashwin~Kumar Rajagopalan,
  Nicholas~Stiles Wilkins, Arvind Rajendran, Ambalavan Jayaraman, and Gokhan
  Alptekin.
\newblock Cycle design and optimization of pressure swing adsorption cycles for
  pre-combustion co2 capture.
\newblock {\em Applied energy}, 254:113624, 2019.

\bibitem{majd2022}
Mahdieh~Mozaffari Majd, Vahid Kordzadeh-Kermani, Vahab Ghalandari, Anis Askari,
  and Mika Sillanp{\"a}{\"a}.
\newblock Adsorption isotherm models: A comprehensive and systematic review
  (2010- 2020).
\newblock {\em Science of The Total Environment}, 812:151334, 2022.

\bibitem{franco2017}
L.~F.~M. Franco, I.~G. Economou, and M.~Castier.
\newblock Statistical mechanical model for adsorption coupled with {SAFT-VR
  Mie} equation of state.
\newblock {\em Langmuir}, 33(42):11291--11298, 2017.

\bibitem{araujo2019}
I.~S. Ara\'ujo and L.~F.~M. Franco.
\newblock A model to predict adsorption of mixtures coupled with {SAFT-VR Mie}
  equation of state.
\newblock {\em Fluid Phase Equilibria}, 496:61--68, 2019.

\bibitem{goncalves2024}
A.~de~F. Gon\c{c}alves, R.~J. Amancio, M.~Castier, and L.~F.~M. Franco.
\newblock Classical density functional theory consistent with the {SAFT-VR Mie}
  equation of state: Development of functionals and application to confined
  fluids.
\newblock {\em Journal of Chemical \& Engineering Data}, 69(10):3645--3659,
  2024.

\bibitem{bardow2025}
V.~Dufour-D\'ecieux, P.~Rehner, J.~Schilling, E.~Moubarak, J.~Gross, and
  A.~Bardow.
\newblock Classical density functional theory as a fast and accurate method for
  adsorption property prediction of porous materials.
\newblock {\em AIChE Journal}, page e18779, 2025.

\bibitem{alessandro2010}
Deanna~M D'Alessandro, Berend Smit, and Jeffrey~R Long.
\newblock Carbon dioxide capture: prospects for new materials.
\newblock {\em Angewandte Chemie International Edition}, 49(35):6058--6082,
  2010.

\bibitem{bui2018}
Mai Bui, Claire~S Adjiman, Andr{\'e} Bardow, Edward~J Anthony, Andy Boston,
  Solomon Brown, Paul~S Fennell, Sabine Fuss, Amparo Galindo, Leigh~A Hackett,
  et~al.
\newblock Carbon capture and storage (ccs): the way forward.
\newblock {\em Energy \& Environmental Science}, 11(5):1062--1176, 2018.

\bibitem{paltsev2021}
Sergey Paltsev, Andrei Sokolov, Xiang Gao, and Martin Haigh.
\newblock Meeting the goals of the paris agreement: temperature implications of
  the shell sky scenario.
\newblock In {\em World Scientific Encyclopedia of Climate Change: Case Studies
  of Climate Risk, Action, and Opportunity Volume 1}, pages 333--339. World
  Scientific, 2021.

\bibitem{choi2009}
Sunho Choi, Jeffrey~H Drese, and Christopher~W Jones.
\newblock Adsorbent materials for carbon dioxide capture from large
  anthropogenic point sources.
\newblock {\em ChemSusChem: Chemistry \& Sustainability Energy \& Materials},
  2(9):796--854, 2009.

\bibitem{verstreken2024}
Margot~FK Verstreken, Nicolas Chanut, Yann Magnin, H{\'e}ctor Octavio~Rubiera
  Landa, Joeri~FM Denayer, Gino~V Baron, and Rob Ameloot.
\newblock Mind the gap: The role of mass transfer in shaped nanoporous
  adsorbents for carbon dioxide capture.
\newblock {\em Journal of the American Chemical Society}, 146(34):23633--23648,
  2024.

\bibitem{rouquerol1994}
J~Rouquerol, D~Avnir, CW~Fairbridge, DH~Everett, JM~Haynes, N~Pernicone, JDF
  Ramsay, KSW Sing, and KK~Unger.
\newblock Recommendations for the characterization of porous solids (technical
  report).
\newblock {\em Pure and Applied Chemistry}, 66(8):1739--1758, 1994.

\bibitem{grossmann2025}
Quirin Grossmann and Marco Mazzotti.
\newblock Mass transfer of co2 in amine-functionalized structured contactors in
  ultra-dilute conditions.
\newblock {\em Industrial \& Engineering Chemistry Research}, 2025.

\bibitem{lin2021}
Jian-Bin Lin, Tai~TT Nguyen, Ramanathan Vaidhyanathan, Jake Burner, Jared~M
  Taylor, Hana Durekova, Farid Akhtar, Roger~K Mah, Omid Ghaffari-Nik, Stefan
  Marx, et~al.
\newblock A scalable metal-organic framework as a durable physisorbent for
  carbon dioxide capture.
\newblock {\em Science}, 374(6574):1464--1469, 2021.

\bibitem{chen2014}
Chao Chen, Dong-Wha Park, and Wha-Seung Ahn.
\newblock Co2 capture using zeolite 13x prepared from bentonite.
\newblock {\em Applied Surface Science}, 292:63--67, 2014.

\bibitem{dhoke2021}
Chaitanya Dhoke, Abdelghafour Zaabout, Schalk Cloete, and Shahriar Amini.
\newblock Review on reactor configurations for adsorption-based co2 capture.
\newblock {\em Industrial \& Engineering Chemistry Research},
  60(10):3779--3798, 2021.

\bibitem{hovington2022}
Pierre Hovington, Omid Ghaffari-Nik, Laurent Mariac, Andrew Liu, Brett Henkel,
  and Stefan Marx.
\newblock Rapid cycle temperature swing adsorption process using solid
  structured sorbent for co2 capture from cement flue gas.
\newblock In {\em Proceedings of the 16th Greenhouse Gas Control Technologies
  Conference (GHGT-16)}, pages 23--24, 2022.

\bibitem{nguyen2022}
Tai~TT Nguyen, Jian-Bin Lin, George~KH Shimizu, and Arvind Rajendran.
\newblock Separation of co2 and n2 on a hydrophobic metal organic framework
  calf-20.
\newblock {\em Chemical Engineering Journal}, 442:136263, 2022.

\bibitem{ymagnin2023}
Yann Magnin, Estelle Dirand, Guillaume Maurin, and Philip~L Llewellyn.
\newblock Abnormal co2 and h2o diffusion in calf-20 (zn) metal--organic
  framework: Fundamental understanding of co2 capture.
\newblock {\em ACS Applied Nano Materials}, 6(21):19963--19971, 2023.

\bibitem{nguyen2023}
Tai~TT Nguyen, George~KH Shimizu, and Arvind Rajendran.
\newblock Co2/n2 separation by vacuum swing adsorption using a metal--organic
  framework, calf-20: Multi-objective optimization and experimental validation.
\newblock {\em Chemical Engineering Journal}, 452:139550, 2023.

\bibitem{chen2023}
Zhihengyu Chen, Ching-Hwa Ho, Xiaoliang Wang, Simon~M Vornholt, Thomas~M
  Rayder, Timur Islamoglu, Omar~K Farha, Francesco Paesani, and Karena~W
  Chapman.
\newblock Humidity-responsive polymorphism in calf-20: A resilient mof
  physisorbent for co2 capture.
\newblock {\em ACS Materials Letters}, 5(11):2942--2947, 2023.

\bibitem{naskar2023}
Supriyo Naskar, Dong Fan, Aziz Ghoufi, and Guillaume Maurin.
\newblock Microscopic insight into the shaping of mofs and its impact on co 2
  capture performance.
\newblock {\em Chemical Science}, 14(38):10435--10445, 2023.

\bibitem{ho2023}
Ching-Hwa Ho and Francesco Paesani.
\newblock Elucidating the competitive adsorption of h2o and co2 in calf-20: new
  insights for enhanced carbon capture metal--organic frameworks.
\newblock {\em ACS Applied Materials \& Interfaces}, 15(41):48287--48295, 2023.

\bibitem{gopalsamy2024}
Karuppasamy Gopalsamy, Dong Fan, Supriyo Naskar, Yann Magnin, and Guillaume
  Maurin.
\newblock Engineering of an isoreticular series of calf-20 metal--organic
  frameworks for co2 capture.
\newblock {\em ACS Applied Engineering Materials}, 2(1):96--103, 2024.

\bibitem{rajendran2024}
Arvind Rajendran, George~KH Shimizu, and Tom~K Woo.
\newblock The challenge of water competition in physical adsorption of co2 by
  porous solids for carbon capture applications--a short perspective.
\newblock {\em Advanced Materials}, 36(12):2301730, 2024.

\bibitem{wang2024}
Xiaoliang Wang, Maytham Alzayer, Arthur~J Shih, Saptasree Bose, Haomiao Xie,
  Simon~M Vornholt, Christos~D Malliakas, Hussain Alhashem, Faramarz Joodaki,
  Sammer Marzouk, et~al.
\newblock Tailoring hydrophobicity and pore environment in physisorbents for
  improved carbon dioxide capture under high humidity.
\newblock {\em Journal of the American Chemical Society}, 2024.

\bibitem{fan2024}
Dong Fan, Supriyo Naskar, and Guillaume Maurin.
\newblock Unconventional mechanical and thermal behaviours of mof calf-20.
\newblock {\em Nature Communications}, 15(1):3251, 2024.

\bibitem{oktavian2024}
Rama Oktavian, Ruben Goeminne, Lawson~T Glasby, Ping Song, Racheal Huynh,
  Omid~Taheri Qazvini, Omid Ghaffari-Nik, Nima Masoumifard, Joan~L Cordiner,
  Pierre Hovington, et~al.
\newblock Gas adsorption and framework flexibility of calf-20 explored via
  experiments and simulations.
\newblock {\em Nature Communications}, 15(1):3898, 2024.

\bibitem{hasting2024}
Jon Hastings, Thomas Lassitter, Nicholas Fylstra, George K.~H. Shimizu, and
  T.~Grant Glover.
\newblock Steam isotherms, co2/h2o mixed-gas isotherms, and single-component
  co2 and h2o diffusion rates in calf-20.
\newblock {\em Industrial \& Engineering Chemistry Research}, 0(0):null, 2024.

\bibitem{krishna2024}
Rajamani Krishna and Jasper~M van Baten.
\newblock Elucidating the failure of the ideal adsorbed solution theory for
  co2/h2o mixture adsorption in calf-20.
\newblock {\em Separation and Purification Technology}, page 128269, 2024.

\bibitem{drwkeska2024}
Joanna Drweska, Filip Formalik, Kornel Roztocki, Randall~Q Snurr, Leonard~J
  Barbour, and Agnieszka~M Janiak.
\newblock Unveiling temperature-induced structural phase transformations and
  co2 binding sites in calf-20.
\newblock {\em Inorganic chemistry}, 63(41):19277--19286, 2024.

\bibitem{lassitter2024}
Thomas Lassitter, Jon Hastings, Nicholas Fylstra, Marc~R Birtwistle, Nikita
  Hanikel, Omar~M Yaghi, George~KH Shimizu, and T~Grant Glover.
\newblock Impacts of adsorbed n2 on the diffusion of co2 and h2o in calf-20
  assessed via a frequency response method.
\newblock 2024.

\bibitem{drwkeska2025}
Joanna Drweska, Kornel Roztocki, and Agnieszka~M Janiak.
\newblock Advances in chemistry of calf-20, a metal--organic framework for
  industrial gas applications.
\newblock {\em Chemical Communications}, 61(6):1032--1047, 2025.

\bibitem{attallah2025}
Ahmed~G Attallah, Volodymyr Bon, Eric Hirschmann, Maik Butterling, Andreas
  Wagner, Rados{\l}aw Zaleski, and Stefan Kaskel.
\newblock Uncovering the dynamic co2 gas uptake behavior of calf-20 (zn) under
  varying conditions via positronium lifetime analysis.
\newblock {\em Small}, 21(14):2500544, 2025.

\bibitem{pereira2025}
Daniel Pereira, Mariana Sardo, Ricardo Vieira, Ildefonso Mar{\'\i}n-Montesinos,
  and Lu{\'\i}s Mafra.
\newblock Enhancing co2 capture via fast microwave-assisted synthesis of the
  calf-20 metal--organic framework.
\newblock {\em Inorganic Chemistry}, 2025.

\bibitem{constant2025}
Noelie Constant, Gwyneth Liske, Shanmuk~Srinivas Ravuru, Anjana Puliyanda,
  Veronique Pugnet, Alejandro Orsikowsky~Sanchez, Sayali~Ramdas Chavan, Philip
  Llewellyn, James~A Sawada, and Arvind Rajendran.
\newblock Binary co2/h2o adsorption on co2 capture metal--organic frameworks
  calf-20, al-fumarate and cau-10-h using microscale dynamic column
  breakthrough.
\newblock {\em Industrial \& Engineering Chemistry Research}, 64(3):1712--1729,
  2025.

\bibitem{duplessis2025}
Julia Duplessis-Kergomard, Meryem Saidi, Olinda Gimello, Bernard Fraisse,
  Fabrice Salles, and Philippe Trens.
\newblock Abnormal adsorption properties of the mof calf-20 as revealed by
  water and methanol vapor sorption.
\newblock {\em Microporous and Mesoporous Materials}, page 113525, 2025.

\bibitem{dilipkumar2025}
Akhil Dilipkumar, Anshu Shukla, Dan Zhao, and Shamsuzzaman Farooq.
\newblock Mixture equilibrium and kinetics of flue gas components in calf-20.
\newblock {\em Chemical Engineering Science}, page 121663, 2025.

\bibitem{dilipkumar2025_2}
Akhil Dilipkumar, Anshu Shukla, Dan Zhao, and Shamsuzzaman Farooq.
\newblock Adsorption equilibrium and transport of co2, n2, and h2o in calf-20.
\newblock {\em Langmuir}, 0(0):null, 0.

\bibitem{vuong1996}
T~Vuong and PA~Monson.
\newblock Monte carlo simulation studies of heats of adsorption in
  heterogeneous solids.
\newblock {\em Langmuir}, 12(22):5425--5432, 1996.

\bibitem{rouquerol2013}
Jean Rouquerol, Fran{\c{c}}oise Rouquerol, Philip Llewellyn, Guillaume Maurin,
  and Kenneth Sing.
\newblock {\em Adsorption by powders and porous solids: principles, methodology
  and applications}.
\newblock Academic press, 2013.

\bibitem{langmuir1918}
Irving Langmuir.
\newblock The adsorption of gases on plane surfaces of glass, mica and
  platinum.
\newblock {\em Journal of the American Chemical society}, 40(9):1361--1403,
  1918.

\bibitem{hill1986}
Terrell~L Hill.
\newblock {\em An introduction to statistical thermodynamics}.
\newblock Courier Corporation, 1986.

\bibitem{nguyen2021}
TTT Nguyen.
\newblock Humid post-combustion co2 capture by vacuum swing adsorption using
  calf-20. 2021.
\newblock {\em There is no corresponding record for this reference}.

\bibitem{myers1983}
AL~Myers.
\newblock Activity coefficients of mixtures adsorbed on heterogeneous surfaces.
\newblock {\em AIChE journal}, 29(4):691--693, 1983.

\bibitem{stanley1993}
Brett~J Stanley, Stephen~E Bialkowski, and David~B Marshall.
\newblock Analysis of first-order rate constant spectra with regularized
  least-squares and expectation maximization. 1. theory and numerical
  characterization.
\newblock {\em Analytical Chemistry}, 65(3):259--267, 1993.

\bibitem{gritti2003}
Fabrice Gritti, Gustaf Gotmar, Brett~J Stanley, and Georges Guiochon.
\newblock Determination of single component isotherms and affinity energy
  distribution by chromatography.
\newblock {\em Journal of Chromatography A}, 988(2):185--203, 2003.

\bibitem{beerdsen2004}
E~Beerdsen, B~Smit, and D~Dubbeldam.
\newblock Molecular simulation of loading dependent slow diffusion in confined
  systems.
\newblock {\em Physical review letters}, 93(24):248301, 2004.

\bibitem{mo1991}
YW~Mo, J~Kleiner, MB~Webb, and MG~Lagally.
\newblock Activation energy for surface diffusion of si on si (001): A
  scanning-tunneling-microscopy study.
\newblock {\em Physical review letters}, 66(15):1998, 1991.

\bibitem{schlaich2025}
Alexander Schlaich, Jean-Louis Barrat, and Benoit Coasne.
\newblock Theory and modeling of transport for simple fluids in nanoporous
  materials: From microscopic to coarse-grained descriptions.
\newblock {\em Chemical Reviews}, 2025.

\bibitem{camp2016}
Jeffrey~S Camp and David~S Sholl.
\newblock Transition state theory methods to measure diffusion in flexible
  nanoporous materials: application to a porous organic cage crystal.
\newblock {\em The Journal of Physical Chemistry C}, 120(2):1110--1120, 2016.

\bibitem{dubbeldam2005}
David Dubbeldam, Edith Beerdsen, Thijs~JH Vlugt, and Berend Smit.
\newblock Molecular simulation of loading-dependent diffusion in nanoporous
  materials using extended dynamically corrected transition state theory.
\newblock {\em The Journal of chemical physics}, 122(22), 2005.

\bibitem{chandler1978}
David Chandler.
\newblock Statistical mechanics of isomerization dynamics in liquids and the
  transition state approximation.
\newblock {\em The Journal of Chemical Physics}, 68(6):2959--2970, 1978.

\bibitem{karger2012}
J{\"o}rg K{\"a}rger, Douglas~Morris Ruthven, Doros~Nicolas Theodorou, et~al.
\newblock {\em Diffusion in nanoporous materials}, volume~48.
\newblock Wiley Online Library, 2012.

\bibitem{dubbeldam2016}
David Dubbeldam, Sof{\'\i}a Calero, Donald~E Ellis, and Randall~Q Snurr.
\newblock Raspa: molecular simulation software for adsorption and diffusion in
  flexible nanoporous materials.
\newblock {\em Molecular Simulation}, 42(2):81--101, 2016.

\bibitem{thompson2022}
Aidan~P Thompson, H~Metin Aktulga, Richard Berger, Dan~S Bolintineanu,
  W~Michael Brown, Paul~S Crozier, Pieter~J In't~Veld, Axel Kohlmeyer, Stan~G
  Moore, Trung~Dac Nguyen, et~al.
\newblock Lammps-a flexible simulation tool for particle-based materials
  modeling at the atomic, meso, and continuum scales.
\newblock {\em Computer Physics Communications}, 271:108171, 2022.

\bibitem{mayo1990}
Stephen~L Mayo, Barry~D Olafson, and William~A Goddard.
\newblock Dreiding: a generic force field for molecular simulations.
\newblock {\em Journal of Physical chemistry}, 94(26):8897--8909, 1990.

\bibitem{campana2009}
Carlos Campa{\~n}{\'a}, Bastien Mussard, and Tom~K Woo.
\newblock Electrostatic potential derived atomic charges for periodic systems
  using a modified error functional.
\newblock {\em Journal of Chemical Theory and Computation}, 5(10):2866--2878,
  2009.

\bibitem{garcia2009}
Almudena Garcia-Sanchez, Conchi~O Ania, Jos{\'e}~B Parra, David Dubbeldam,
  Thijs~JH Vlugt, Rajamani Krishna, and Sofia Calero.
\newblock Transferable force field for carbon dioxide adsorption in zeolites.
\newblock {\em The Journal of Physical Chemistry C}, 113(20):8814--8820, 2009.

\bibitem{hockney2021}
Roger~W Hockney and James~W Eastwood.
\newblock {\em Computer simulation using particles}.
\newblock crc Press, 2021.

\bibitem{peng1976}
Ding-Yu Peng and Donald~B Robinson.
\newblock A new two-constant equation of state.
\newblock {\em Industrial \& Engineering Chemistry Fundamentals}, 15(1):59--64,
  1976.

\bibitem{stanley_guiochon1993}
Brett~J. Stanley and Georges Guiochon.
\newblock Numerical estimation of adsorption energy distributions from
  adsorption isotherm data with the expectation-maximization method.
\newblock {\em The Journal of Physical Chemistry}, 97(30):8098--8104, 1993.

\bibitem{nist}
{National Institute of Standards and Technology}.
\newblock Thermophysical properties of fluid systems - nist chemistry webbook,
  srd 69.
\newblock Technical report, U.S. Department of Commerce, Washington, D.C.,
  2023.
\newblock [Online; Available from
  \url{https://webbook.nist.gov/cgi/cbook.cgi?ID=C124389}; Accessed
  24-March-2025].

\end{thebibliography}

\end{document}